\documentclass{sn-jnl}
\usepackage{graphicx}%
\usepackage{multirow}%
\usepackage{amsmath,amssymb,amsfonts}%
\usepackage{mathrsfs}%
\usepackage[title]{appendix}%
\usepackage[numbers,sort&compress]{natbib} % Q: need to see if it is what we want, without it the references were overflowing on the margins
\usepackage{xcolor}%
\usepackage{textcomp}%
\usepackage{manyfoot}%
\usepackage{booktabs}%
\usepackage{algorithm}%
\usepackage{algorithmicx}%
\usepackage{algpseudocode}%
\usepackage{listings}%
\usepackage[utf8]{inputenc}
\usepackage{subcaption}
\usepackage{dirtytalk}
\usepackage{indentfirst}
\usepackage{paralist}
\usepackage{svg}
\usepackage{multicol}
\usepackage{tabularx} % for the X column type
\usepackage{pdfpages}
\usepackage{booktabs}
\usepackage{graphicx}
\usepackage{subcaption}
\usepackage{array}         
\usepackage{tikz}        
\usepackage{xcolor}  
\usepackage{cleveref}
\usepackage{csvsimple}
\usepackage{wasysym}
\usepackage[inline]{enumitem}

\lstdefinestyle{javaCustom}{
  language=Java,
  basicstyle=\ttfamily\small,
  keywordstyle=\color{blue}\bfseries,
  commentstyle=\color{gray}\itshape,
  stringstyle=\color{orange},
  numbers=left,
  numberstyle=\tiny\color{gray},
  stepnumber=1,
  numbersep=5pt,
  backgroundcolor=\color{white},
  frame=single,
  tabsize=2,
  breaklines=true,
  captionpos=b
}

\AtBeginDocument{\DeclareCaptionSubType{lstlisting}}
\crefname{sublstlisting}{listing}{listings}
\Crefname{sublstlisting}{Listing}{Listings}

\usepackage{xspace}
\usepackage{xfrac}
\newcommand{\gptidea}{$\mathrm{GPT_{IDEA}}$\xspace}
\newcommand{\pidea}{$\mathrm{P_{IDEA}}$\xspace}
\newcommand{\pimpl}{$P_{IMPL}$\xspace}
\newcommand{\gptimpl}{$\mathrm{GPT_{IMPL}}$\xspace}

\newcommand{\gptideagptimpl}{\boldmath\gptidea{}\gptimpl}
\newcommand{\cone}{\boldmath\gptidea{}\gptimpl}
\newcommand{\gptideapimpl}{\boldmath\gptidea{}\pimpl}
\newcommand{\ctwo}{\boldmath\gptidea{}\pimpl}
\newcommand{\pideagptimpl}{\boldmath\pidea{}\gptimpl}
\newcommand{\cthree}{\boldmath\pidea{}\gptimpl}
\newcommand{\pideapimpl}{\boldmath\pidea{}\pimpl}
\newcommand{\cfour}{\boldmath\pidea{}\pimpl}

\begin{document}

\title{How Students Use Generative AI for Software Testing: An Observational Study}


\author*[1]{\fnm{Baris} \sur{Ardic}}\email{b.ardic@tudelft.nl}

\author[1]{\fnm{Quentin} \sur{Le Dilavrec}}\email{q.ledilavrec@tudelft.nl}

\author[1]{\fnm{Andy} \sur{Zaidman}}\email{a.e.zaidman@tudelft.nl}

\affil*[1]{\orgdiv{Computer Science}, \orgname{Delft University of Technology}, \orgaddress{\city{Delft}, \postcode{2628XE}, \state{State}, 
\country{Netherlands}}}

%\author{Baris Ardic \and Quentin Le Dilavrec \and Andy Zaidman}

%\institute{
%Baris Ardic \and Quentin Le Dilavrec \and Andy Zaidman \at
%Delft University of Technology, Netherlands \\
%\email{b.ardic@tudelft.nl, q.ledilavrec@tudelft.nl, %a.e.zaidman@tudelft.nl}
%}

%\affil[2]{\orgdiv{Department}, \orgname{Organization}, \orgaddress{\street{Street}, \city{City}, \postcode{10587}, \state{State}, \country{Country}}}

%\affil[3]{\orgdiv{Department}, \orgname{Organization}, \orgaddress{\street{Street}, \city{City}, \postcode{610101}, \state{State}, \country{Country}}}


\abstract{The integration of generative AI tools like ChatGPT into software engineering workflows opens up new opportunities to boost productivity in tasks such as unit test engineering. However, these AI-assisted workflows can also significantly alter the developer’s role, raising concerns about control, output quality, and learning, particularly for novice developers.
This study investigates how novice software developers with foundational knowledge in software testing interact with generative AI for engineering unit tests. Our goal is to examine the strategies they use, how heavily they rely on generative AI, and the benefits and challenges they perceive when using generative AI-assisted approaches for test engineering. We conducted an observational study involving 12 undergraduate students who worked with generative AI for unit testing tasks.
We identified four interaction strategies, defined by whether the test idea or the test implementation originated from generative AI or the participant. Additionally, we singled out prompting styles that focused on oneshot or iterative test generation, which often aligned with the broader interaction strategy. Students reported benefits including time-saving, reduced cognitive load, and support for test ideation, but also noted drawbacks such as diminished trust, test quality concerns, and lack of ownership. While strategy and prompting styles influenced workflow dynamics, they did not significantly affect test effectiveness or test code quality as measured by mutation score or test smells.
%In turn, we develop guidelines to help software engineers avoid ineffective practices. 
}

\keywords{software testing, generative AI, human-AI interaction, AI4SE}



\maketitle

\section{Introduction}
\label{section:introduction}
%\section{Student and Generative AI Experiment}

The advances of generative artificial intelligence (generative AI) tools are changing how software is developed~\cite{russo2024generative,hassanFSE2024}. While these AI-assisted workflows enable higher productivity~\cite{weber2024,zieglerCACM2024}, they likely also result in humans being or feeling less directly involved in how software is written~\cite{SchecterCHI2025}. We observe that generative AI becomes increasingly integrated into software engineering workflows, including software testing, where large language models have already been investigated for potential applications to tasks such as unit test generation~\cite{elhajiAST2025,deljouyiICSE2025}, test oracle creation~\cite{molinaTOSEM2025}, debugging~\cite{majdoubESEM2024}, and program repair~\cite{wang2024software}.

Because of the increasing importance of generative AI-assisted software engineering workflows, it is both important and interesting to better understand how developers interact with these tools in practice.
This understanding is fundamental because the effectiveness and reliability of AI-assisted workflows depend not only on the technical capabilities of the tools, but also on how developers use, adapt to, and critically assess their outputs. This is also echoed in the realm of testing, where Ardic et al. have previously called upon the research community to further qualitatively explore how AI tools are used~\cite{ardic2025qualitative}. 

%\andy{Isn't this a bit of a repeat of the previous 2 sentences modulo the novice developer?} As generative AI becomes increasingly integrated into software engineering workflows, understanding how developers, especially novices, interact with these tools is becoming more important. While much of the current literature focuses on performance metrics or output quality, recent work has highlighted the value of qualitative perspectives in software testing, particularly in capturing how people experience and adapt to evolving tools and practices~\cite{ardic2025qualitative}. \baris{That same study also calls upon the research community to further qualitatively explore how AI tools are used in the context of testing~\cite{ardic2025qualitative}}
%A recent systematic mapping study of qualitative research in software testing culminated in a research agenda aimed at expanding qualitative inquiry into underexplored topics, including the use of AI tools in testing contexts \cite{ardic2025qualitative}. This agenda emphasizes the need to better understand how practitioners engage with emerging technologies, especially in settings where human decision-making plays a central role.

%Our participants had already completed a dedicated course on software testing, so our goal was not to teach or assess their understanding of testing concepts. Instead, we examine how students intuitively use generative AI, specifically ChatGPT, when approaching a unit testing task. 
Motivated by this call and the growing relevance of human–AI collaboration in software testing, we set out to explore how students, already equipped with foundational knowledge from a dedicated course on software testing, make use of generative AI, specifically ChatGPT, when faced with a unit testing task. Our objective was not to assess their theoretical understanding of testing concepts, but to investigate how they intuitively incorporate AI into their workflow.

By analyzing their prompting strategies, workflows, and reflections, we aim to uncover the skills that are necessary, and the challenges that arise when AI is integrated into test engineering. In doing so, we contribute to the ongoing research agenda on understanding human–AI interaction in software engineering in general~\cite{choudhuriICSE2024}, and software testing contexts in particular~\cite{ardic2025qualitative}. 

The following research questions guide our investigation. 
%The research questions in the next subsection guide our research:

%Of course, software testing is an essential part of these changes. We aim to explore how novice software testers, particularly students, will interact with generative AI tools during unit testing and to identify which skills and behaviors might enhance their effectiveness in an AI-integrated workflow. In the current plan, we select ChatGPT as our generative AI tool and TU Delft students as our audience.  We aim to answer the following research questions:

%\subsection{Research Questions}

\begin{description}
    %\item \textbf{RQ1:} What strategies do students employ when integrating generative AI into unit testing?
    \item[\textbf{RQ1:}] What strategies do students adopt when incorporating generative AI into unit testing workflows?
\end{description}
An AI-assisted workflow offers both advantages and potential challenges, making it crucial to understand how students approach working with generative AI in a unit test engineering context. Since there is no predefined user guide or a finite set of best practices for using AI in this context, students must independently develop their workflows. By investigating the strategies they adopt, we can identify both effective practices and common missteps. This insight allows us to encourage behaviors that maximize the benefits of AI while raising awareness of potential risks. %Furthermore, understanding how students navigate AI-assisted test engineering can inform training programs that help individuals make the most of this technology. It can also contribute to refining future AI tools, ensuring they better support developers in software testing tasks.
\begin{description}
\item[\textbf{RQ2:}] How do students prompt generative AI?
\end{description}

Examining how students formulate their prompts provides valuable insight into how they use generative AI for unit testing. 
%There is no predefined set of use cases for generative AI for test engineering.
For example, this can help us uncover the specific use cases they pursued, whether they used generative AI to generate complete test cases, refine the cases they implemented themselves, or improve their understanding of testing concepts.
%Additionally, by tracking prompts throughout the course of an assignment, we can determine whether students’ use of AI evolves over time. Do they adjust their prompting strategies as they become more familiar with the tool, or do they remain consistent in their approach? 
%Understanding these patterns helps us gauge how students engage with AI in a real-world testing scenario and what aspects of AI-assisted workflows they find most beneficial.
Analyzing these use patterns helps us understand how students engage with AI in practical testing scenarios and which aspects of AI integration they find most useful.

\begin{description}
\item[\textbf{RQ3:}] What are the benefits and challenges of a generative AI-assisted test workflow for students?
\end{description}
AI tools promise to improve productivity, but it is essential to evaluate whether students actually experience tangible benefits in terms of efficiency, learning, and software quality. Do they find AI-generated test cases useful? Does AI assistance help them better understand unit testing principles, or does it make them overly reliant on automation? To answer these questions, we examine both the perceived benefits reported by students and the benefits observed by us, researchers. While students can provide insight into how AI affects their workflow, confidence, and learning experience, researchers can analyze whether AI genuinely enhances the quality of their test cases and improves their approach to unit testing. %This distinction is crucial in determining whether AI tools are genuinely enhancing the process and whether their adoption in testing workflows leads to meaningful improvements in both software quality and the human experience of the process.

%\begin{itemize}
%\item \textbf{RQ4:} \baris{will take this out }What common mistakes or oversights do students make when using generative AI for unit testing?
%\end{itemize}
%While AI can accelerate test creation, it may also introduce new challenges and risks, such as generating incomplete or incorrect test cases that students might fail to recognize. Investigating the common mistakes that students make—whether in interpreting AI outputs, verifying test adequacy, or refining test cases—will help pinpoint gaps in knowledge and training. This question is critical because it highlights the potential risks of AI reliance in software testing and informs the development of best practices and educational interventions to ensure students remain engaged as thoughtful, critical evaluators rather than passive consumers of AI-generated tests.
%\hfill \break

%We drew on four sources of data for our analysis: think-aloud protocols, observation notes, video recordings of the assignment sessions, and interview transcripts. Using open coding followed by thematic analysis, we qualitatively analyzed this data to identify patterns in how students engaged with generative AI in the context of unit testing. Our main findings are:
%\baris{Here I kind of wrote myself into talking about results, so I stopped here}

To address these research questions, we designed an observational study with 12 students who engaged in two consecutive unit testing tasks, during which they were free to use ChatGPT as they saw fit. Our goal was not to assess their theoretical knowledge of testing concepts, since all participants had previously completed a foundational course on software testing, but to observe how they integrated AI into their workflows.

%During the sessions, participants were encouraged to think aloud while working through two unit testing assignments. We recorded their screen activity, prompts, and outputs, and conducted post-task interviews to capture their perceptions and decision-making processes. This mixed-source dataset enabled a fine-grained analysis of their interactions with generative AI.

Our study makes three main contributions:
\begin{itemize}
    \item We identify the test engineering strategies students employ when using generative AI and categorize them into four distinct types.
    \item We analyze the prompting behaviors of students across assignments, uncovering patterns in how they formulate requests and structure their generative AI collaboration.
    \item We examine both the perceived and observed benefits and challenges of using generative AI for software testing, including impacts on workflow, trust, and learning.
\end{itemize}

The remainder of the paper is structured as follows. Section~\ref{section:background} reviews related work on generative AI in software engineering and education. 
Section~\ref{section:methodology} outlines our methodology, including experimental setup, data collection, and data analysis process. Section~\ref{section:results} presents our results, organized around the three research questions. Section~\ref{section:discussion} discusses key insights and implications. Section~\ref{section:threats} reflects on threats to validity, and Section~\ref{section:conclusion} concludes the paper.

\section{Background and Related Work}\label{section:background}

Recent research on large language models (LLMs) in software engineering education has primarily focused on two areas: evaluating their effectiveness in technical tasks and examining their use in educational settings. We discuss both topics in the subsequent subsections.

\subsection{LLMs for Software Engineering Tasks}

The performance and usability of ChatGPT is being considered for many use cases in software engineering. White et al.~\cite{white2023prompt} proposed a catalog of prompt engineering patterns to guide ChatGPT in tasks such as requirements elicitation and code refactoring. Their structured prompting strategies aim to improve the consistency, clarity, and maintainability of generated outputs.

%Several studies have investigated the capabilities and limitations of LLMs in software testing and engineering contexts. 
Jalil et al.~\cite{jalil2023chatgpt} evaluated ChatGPT’s performance on software testing questions, finding that it produced correct or partially correct answers 55.6\% of the time, with explanations reaching similar accuracy. The study also showed that contextual prompting improved performance, although the model’s self-reported confidence was not a reliable indicator of correctness.

For LLM applications in test generation, Jain and Le Goues~\cite{jain2025testforge} introduced \textit{TestForge}, an agentic framework that iteratively refines test suites using feedback from code execution and coverage reports. Starting from zero-shot outputs, the system dynamically updates test cases to improve performance and cost-efficiency. On the TestGenEval benchmark, TestForge achieved higher accuracy and coverage than both classical genetic programming tools and non-agentic LLM baselines, demonstrating the value of feedback-driven refinement in automated test generation.

Mock et al.~\cite{mock2024generative} explored generative AI use in test-driven development. Their initial findings report that generative AI can accelerate the process and reduce effort, while cautioning that use without active developer oversight may lead to brittle or misleading code, particularly for less experienced developers.

\subsection{Use in Educational Contexts}

Other studies have explored the integration of generative AI tools into software engineering education. Haldar et al.~\cite{haldar2025exploring} examined how postgraduate students used ChatGPT and GitHub Copilot to generate testing artifacts. While students found the tools helpful for speeding up their work, authors emphasized the continued need for manual oversight to ensure quality.

Similarly, Mezzaro et al.~\cite{mezzaro2024empirical} observed that students who relied heavily on a GPT-based assistant within a gamified platform produced fewer effective test cases and demonstrated weaker learning outcomes. These results underscore the importance of guided use.

Daun and Brings~\cite{daun2023chatgpt} took a broader view, arguing that the rise of generative AI calls for changes to both teaching practices and curricula. They recommended shifting the focus from coding to design and critical thinking, and emphasized the need for supervised use of AI tools to avoid reinforcing misconceptions.

Rahe and Maalej~\cite{rahe2025programming} conducted an in-depth study on how undergraduate programming students use ChatGPT during a programming exercise. Their findings reveal that many students, especially those with prior GenAI experience, gravitate toward using ChatGPT for direct code generation rather than for support or conceptual understanding. A common pattern observed was a cycle of requesting code, submitting the output, encountering errors, and re-prompting without critical reflection—a strategy that often proved ineffective. The study also highlights how overreliance on GenAI can lead to diminished autonomy and problem-solving agency, raising concerns for software engineering education.

Choudhuri et al.~\cite{choudhuriICSE2024} evaluated ChatGPT’s role in supporting students with software engineering tasks through a controlled study. They found no improvement in productivity or self-efficacy, but significantly higher frustration among students using ChatGPT. The study identified recurring issues such as hallucinated responses and incomplete assistance, which were linked to violations of Human AI interaction guidelines, highlighting the need for better alignment with educational goals.

\subsection{Qualitative Methods}

In studying how students interact with generative AI during software testing tasks, qualitative methods offer unique advantages. While quantitative metrics such as test coverage or prompt counts can reveal patterns in output, they often fail to capture the nuanced behaviors, decision-making processes, and mental models that underlie those patterns. Understanding why and how students choose to engage with AI, what challenges they encounter, and how they adapt their strategies requires close attention to their experiences, actions, and reflections.

Qualitative inquiry is particularly valuable in emerging contexts like AI-assisted software testing, where user practices are not yet well-established. In such settings, it enables researchers to explore open-ended questions and inductively surface patterns in human–AI interaction that might otherwise be overlooked. This aligns with recent reflections by Treude and Storey~\cite{treude2025generative}, who argue that the blurring of socio-technical boundaries introduced by AI tools complicates efforts to isolate human behavior from AI-mediated processes. They emphasize that qualitative methods must evolve to study AI not only as a tool but as an active participant in developer workflows.

Trinkenreich et al.~\cite{trinkenreich2025get} similarly underscore the importance of qualitative approaches in making sense of complex and interrelated findings across multiple data sources, particularly when studying how developers experience, interpret, and adapt to the use of LLMs. They argue that preserving human agency and interpretability is essential for upholding scientific rigor, especially when studying the influence of AI on developer behavior, team dynamics, and socio-technical interactions.

To address the methodological challenges identified in prior work, our study design incorporates multiple qualitative data sources: think-aloud protocols, screen recordings, observation notes, and post-task interviews. These sources allow us to trace both observable behavior and participants’ internal reasoning. By capturing student actions in real time and following up with reflective interviews, we mitigate the risk of overlooking the human experience behind AI-mediated processes, as cautioned by Treude and Storey~\cite{treude2025generative}. Furthermore, our use of open and axial coding, followed by thematic analysis, supports the kind of interpretive rigor advocated by Trinkenreich et al.~\cite{trinkenreich2025get}, enabling us to study AI as an active agent in student workflows while preserving human agency and interpretability in the analysis process.

To meet the methodological needs highlighted in prior work, our study combines think-aloud protocols, screen recordings, observation notes, and interviews to capture both observable behavior and participants’ reasoning. This multi-source design helps mitigate the risk of overlooking human experiences in AI-mediated workflows~\cite{treude2025generative}, while our extensive qualitative data collection and analysis aim to preserve human agency and interpretive depth throughout the research process~\cite{trinkenreich2025get}.

\section{Methodology}
\label{section:methodology}

We conducted an observational study to examine how students interact with generative AI during unit testing, capturing both behavioral data and participant reflections through screen recordings, think-aloud protocols, and post-task interviews. Section~\ref{ExperimentSetup} describes the experimental setup, tasks, and interview design. The remaining subsections describe our data analysis procedures, organized by research question.

\begin{figure}[!thb]
    \centering
   \includegraphics[width=\linewidth]{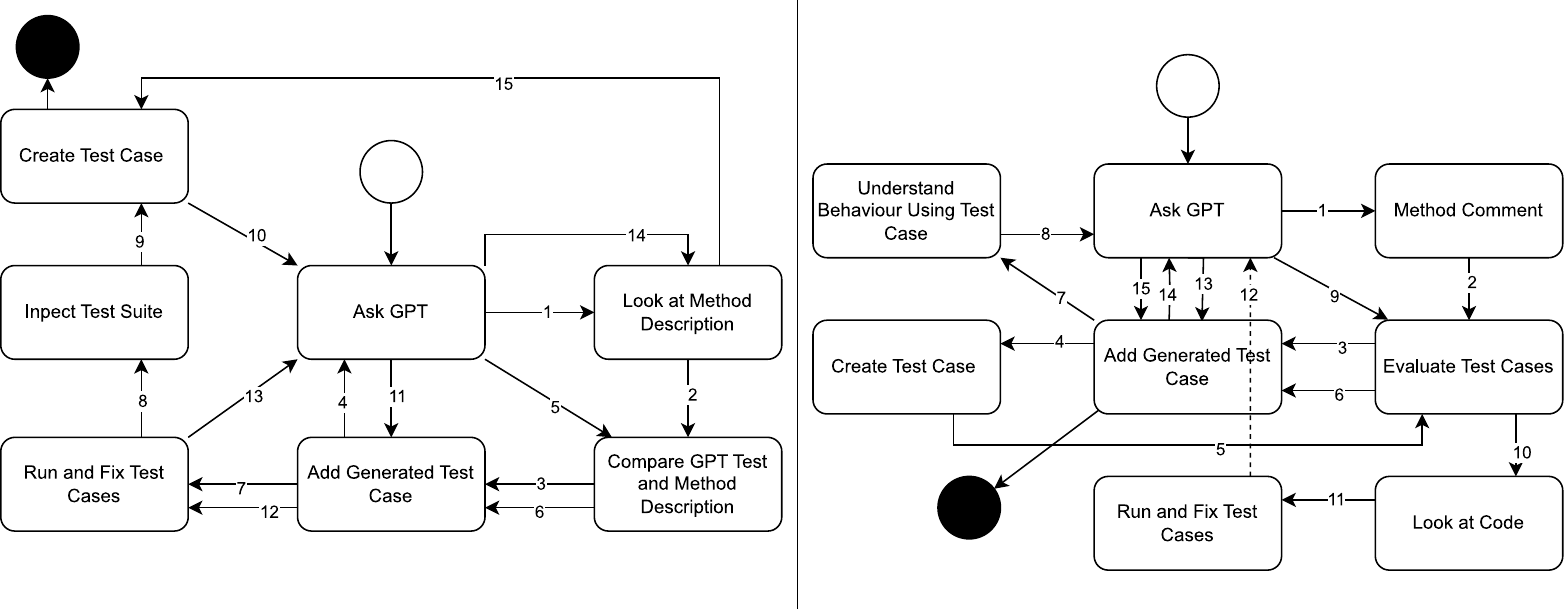}
  \caption{Flow diagram examples for pilot experiment participants}
  \label{fig:pfd}
 
\end{figure}

\subsection{Experimental Setup}\label{ExperimentSetup}
\begin{figure}[!b]
    \centering
   %\includesvg[width=\linewidth]{pictures/"ExperimentFlow".svg}
   \includegraphics[width=\linewidth]{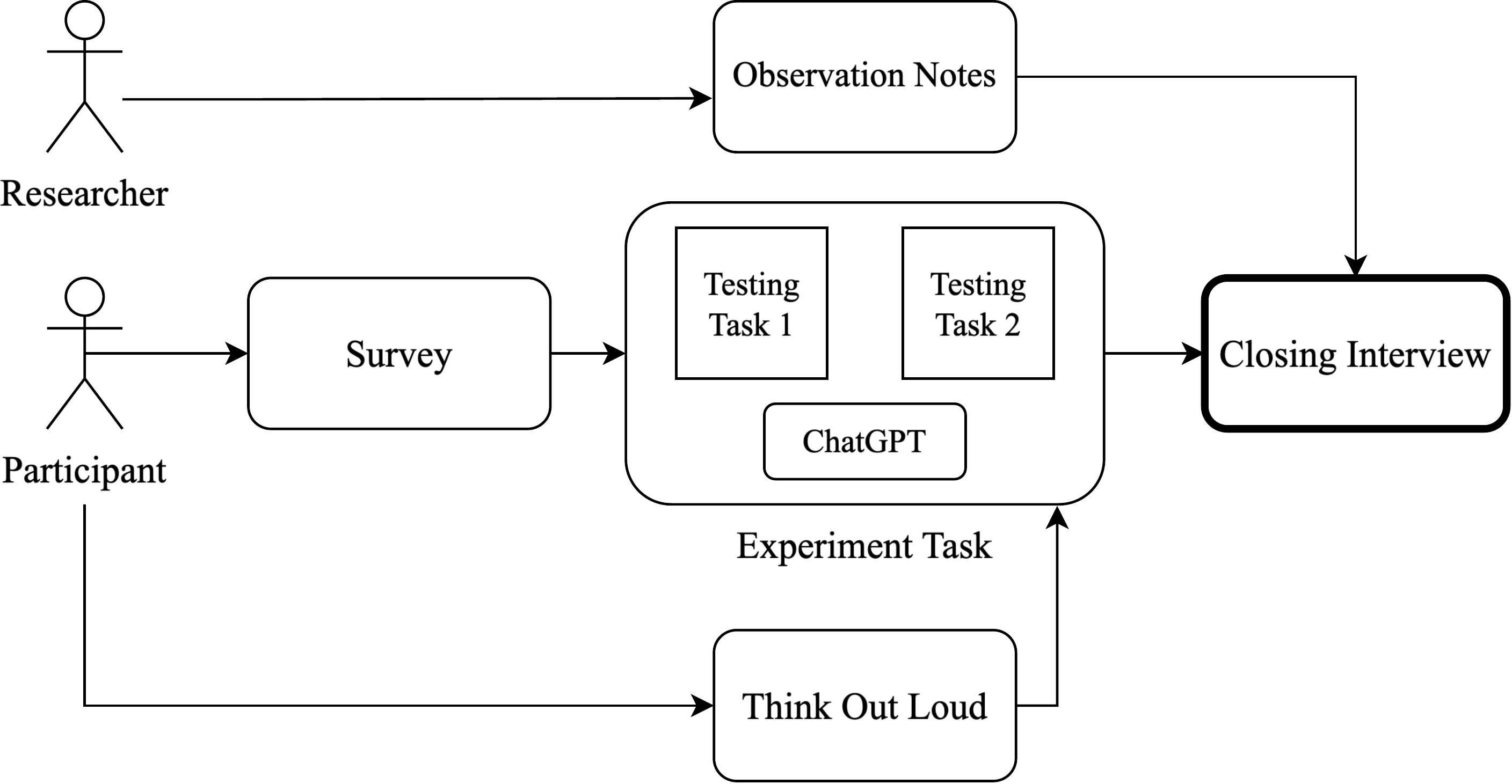}
  \caption{Experiment Flow}
  \label{fig:expflow}
 
\end{figure}

At the start of the experimental session, each participant completed a short questionnaire to provide relevant background information. Subsequently, they proceeded with two consecutive unit testing tasks, during which they were free to use ChatGPT in any way they found helpful. The participants used the free browser-based version of ChatGPT, which at the time was running GPT-3.5. The experiment was conducted over a period spanning from December 2023 to May 2024. 

Throughout the session, we encouraged participants to think out loud, enabling us to capture their reasoning and decision-making processes in real time. In parallel, we recorded observation notes to document the behavior and interactions of the participants as they worked. We also recorded their screen activity, including all prompts and outputs, to enable detailed reconstruction of their workflows.
Once participants indicated that they had completed both tasks, we conducted a post-task interview to reflect on their experiences and explore the strategies they employed. This mixed-source dataset comprising screen recordings, prompt logs, observation notes, and interviews enables a fine-grained analysis of their interactions with generative AI.
The overall structure of the session is summarized in Figure~\ref{fig:expflow}.

The unit testing tasks used in this experiment were originally used in a study by Aniche et al.~\cite{AnicheTSE2022}, which investigated how developers engineer test cases; their study was done without the possibility of making use of generative AI. Reusing these assignments enables us to make some comparisons between our findings and theirs, particularly in terms of code metrics related to the resulting test suites. 

The two tasks were designed around two methods that were taken from Apache Commons Lang~\cite{apacheCommonsLang}\footnote{https://commons.apache.org/lang hash for the commit used:
e0b474c0d015f89a52.}. The two methods were chosen such that they do not depend on any other class, and they are relatively short (between 20 to 30 lines of code) string manipulation methods.
For each task, participants were asked to write unit tests for one of the methods that were devoid of any existing unit tests. The two methods are as follows:

\begin{itemize}
    \item initials(String str, char... delimiters): Extracts the initial
characters from each word in the String. All first
characters after the defined delimiters are returned
as a new string.
    \item substringsBetween(String str, String open, String close):
Searches a String for substrings delimited by a start
and end tag, returning all matching substrings in an
array. 
\end{itemize}

In addition, we adopt a think-aloud approach similar to that used in their study, where participants are asked to verbalize what they are doing and why.
In our instructions to the participants, we advise them to try and answer these questions to help them think out loud:

\begin{itemize}
    \item What are you doing right now?
    \item Why would you test this?
    %\item Why would you not test this?
    \item What challenges are you facing?
    \item What are you not understanding?
    \item What is the next step?
\end{itemize}
Specifically while interacting with ChatGPT, the following questions were relevant to think out loud:
\begin{itemize}
    \item Why are you writing this prompt?
    \item What do you think of the generated answer?
\end{itemize}

The think-aloud approach proved useful in helping us capture and structure observation notes, and also enhanced the value of the video recordings by providing a narrative of the participant’s actions.

We conducted three pilot sessions to refine our experimental design and rehearse data collection procedures. These sessions informed the development of our think-aloud questions and contributed to the formulation of the interview guidelines.

We visualized our observation notes to understand participant behavior better. Figure~\ref{fig:pfd} shows an example with two flow diagrams from the same participant, illustrating how their approach evolved between assignments. In this example, we can observe a change in the strategy from the first assignment (left-hand side) to the second (right-hand side). After experiencing the accuracy of ChatGPT, the participant put it in a more central position in their workflow. We believe these flow diagrams will make our analysis easier to communicate.

We recruited bachelor students from TU Delft on the condition that they had passed the first-year mandatory course on software testing and quality~\cite{aniche2019pragmatic}. Recruitment was conducted via an internal university Mattermost~\cite{mattermost} channel. The sessions were carried out in a 1 on 1 setting where the researchers took observation notes that were later enhanced by the screen recordings of the participants.

This study was reviewed and approved by the Human Research Ethics Committee of Delft University of Technology. Participants gave informed consent to take part in the study and were made aware that all findings would be published in anonymized form, with no personally identifiable information disclosed.

%\subsection{Pilot Experiments}\label{PilotExperiments}

%To solidify the design of the experiment and practice our data collection methods, we conducted three pilot experiment sessions. This process helped with refining the details of the experiment. According to our experiences with the think-aloud method in the pilot experiments, we expect that some participants to have a harder time articulating what they are doing.  Even if that is the case, the think-aloud process helped the researcher take and organize notes. 

%We conducted three pilot sessions to refine our experimental design and rehearse data collection procedures. Based on our experience with the think-aloud method during these sessions, we expect some participants may struggle to articulate their actions. We therefore created the think-aloud guideline questions elaborated in Section~\ref{ExperimentSetup}

\subsubsection{Interview Guide}

To complement our observational data, we conducted post-task interviews with participants. These interviews were semi-structured, allowing us to explore core themes while also following up on interesting or unexpected behaviors observed during the task. In particular, we expanded our questioning when a participant demonstrated a unique approach or mindset that had not been seen up to that point in the data collection process.

The following questions formed the basis of our interview guide:

\begin{itemize}
    \item Do you think you could do this task entirely by yourself without needing any help?
    \item And what has being able to use ChatGPT changed for you?
    \item Do you think working the way you just did changes which skills you use while you are working?
    \item Do you think working the way you did requires a different mindset? 
    \item What do you think about the test suite you now have?
    \item Do you think you can improve it in any way?
\end{itemize}

%\subsection{Data Analysis}\label{DataAnalysis}

\subsection{\textbf{RQ1: What strategies do students adopt when incorporating generative AI into unit testing workflows?}}

For RQ1, we analyzed participant video recordings of the assignments using an open coding approach to identify the actions they performed. The think-aloud protocol further enriched our analysis by revealing participants’ intentions and decision-making processes, which helped us interpret and categorize their actions meaningfully.
%We also measured the duration of these actions to understand how students allocate their time within an AI-assisted workflow. These were later used in discussing points related to productivity.   ]

The first author, who also served as the observer during data collection, conducted the initial open coding~\cite{williams2019art} by detecting and labeling the specific activities performed by participants. In the subsequent axial coding~\cite{williams2019art} stage, we assigned these open codes to broader strategy categories. To assess the reliability of this classification, another author independently assigned the open codes to strategies. We then calculated Cohen’s Kappa~\cite{gisev2013interrater} between the two raters, which yielded a value of $0.75$, indicating substantial agreement. Following this, both raters held a consensus meeting to resolve disagreements and finalize the mapping of actions to strategies.

This coding process resulted in the identification of four core strategies (C1-C4); these strategies describe whether the participant or ChatGPT was the main responsible for ideation and/or implementation. In addition, we identified supporting categories such as documentation use and mental model building. While the detailed definitions of these strategies are presented in Section~\ref{r_strategies}, we used the distribution of coded actions, along with the provenance of generated assertions, to determine each participant’s \textit{dominant strategy} for each assignment. This provided a structured foundation for interpreting the behavior of the participants in later analyses.

\subsubsection{Assertion Matching}\label{section:assertionmatching}
\begin{figure}[tb]
  \setcaptiontype{lstlisting}
    \begin{minipage}[t]{1\textwidth}
        \begin{lstlisting}[style=javaCustom]
String result = YourClassName.initials("HelloWorld");
assertEquals("HW", result);
        \end{lstlisting}
        \vspace{-3mm}
        \subcaption{Initial generated test}
        \label{sublst:vari:0_initial}
        \begin{lstlisting}[style=javaCustom]
assertEquals("HW", YourClassName.initials("HelloWorld"));
        \end{lstlisting}
        \vspace{-3mm}
        \subcaption{Test after asking ChatGPT to merge previously generated test cases into a single one, the inlining action was not explicitly asked by the participant}
        \label{sublst:vari:1_merge}
        \begin{lstlisting}[style=javaCustom]
assertEquals("HW", InitialsUtils.initials("HelloWorld"));
        \end{lstlisting}
        \vspace{-3mm}
        \subcaption{Test after renaming \texttt{YourClassName} to \texttt{InitialsUtils} using ChatGPT}
        \label{sublst:vari:2_rename}
        \begin{lstlisting}[style=javaCustom]
assertEquals("H", InitialsUtils.initials("HelloWorld"));
        \end{lstlisting}
        \vspace{-3mm}
        \subcaption{Final test after a manual correction of the expected value}
        \label{sublst:vari:3_fix_expect}
    \end{minipage}
    \caption{Example of iterative test generation and refinement}
    \label{lst:vari}
\end{figure}

\begin{figure}[tb]
\begin{lstlisting}[style=javaCustom, caption={Example of test produced by one of the participant }]
@Test
void testBasics() {
    // Test with a simple sentence and default delimiter (space)
    String result1 = initials("Hello World");
    assertEquals("HW", result1);

    // Test with custom delimiter (comma)
    String result2 = initials("Alice,Bob,Charlie", ',');
    assertEquals("ABC", result2);
}
\end{lstlisting}
\end{figure}

% restructured methodology of the provenance analysis
% ## goal/motivation
% ## definition of the terms using lst:vari
% ## methodology of provenance analysis
% ### methodology of pattern synth
% ### methodology of provenance analysis given that we extract the relevant elements using the synthetized patterns
% ## restating the goals and stating the expected results in general

To support our analysis of participant strategies and assess ChatGPT’s contribution to the final test suites of participants, we implemented an automated provenance analysis pipeline.
% To differentiate between tests suggested by ChatGPT and tests crated by participants, and determine their dominant strategy, we automated the analysis comparing the ChatGPT conversations to the produced tests.
By automating the comparison of conversations and tests produced by participants, we aim to reproducibly infer when tests were generated by ChatGPT.
Provenance analysis approaches have already been used to compare code artifacts and ChatGPT conversations, notably through text-based approaches~\cite{przymus2024learned}. However, due to the semantic nature of tests and their expected low structural diversity, we use a syntactic representation.{ % How I Learned to Stop Worrying and Love ChatGPT (MSR24) https://dl.acm.org/doi/pdf/10.1145/3643991.36450 Section 2.2 Comparing ChatGPT conversations to code
% Explaining prescisely what we consider a test: structural, assertion- and block- based, exact or similar
% Define conversation in terms of prompts and responses
% Q: yes theirs is very text-based, they use the gestalt approach to pattern matching. With the patterns, our provenance analysis is specialized exactly for comparing test assertions, for examples, we ended up ignoring the expected values (because it does not change the behavior of the tested method) and the exact names of methods (because it is a minor aspect directly catched by the compiler). Analysing additional aspects related to refactorings can be shown but they do not contribute to the current story
}

To illustrate the provenance analysis challenge, we look at \Cref{lst:vari}, which presents tests at different steps of an iterative refinement process as done by one of our participants. The examples present the same test assertion, i.e., an assertion statement and all the other statements necessary to evaluate the assertion.
% We were able to categorize the different test patterns and count their occurrences.
\Cref{sublst:vari:0_initial} is semantically equivalent to Listing~\ref{sublst:vari:1_merge}, but structurally different.
\Cref{sublst:vari:1_merge,sublst:vari:2_rename,sublst:vari:3_fix_expect} have the same syntactic structure, but different references and literals. Notably, the code in Listing~\ref{sublst:vari:2_rename} can be compiled and the test passes. Thus, we choose to consider  \cref{sublst:vari:0_initial,sublst:vari:1_merge,sublst:vari:2_rename} to be equivalent. We consider Listing~\ref{sublst:vari:3_fix_expect} to be similar to the other three.

%\andy{A lot of detail in the following text. What I am missing though: what is the goal and that is to understand/analyze the provenance of assertions. That has not been said up until now. \textbf{And while the text has much improved, this is still true!!!}}% Q: THe approach ended up relying more and more on the patterns to give more prescise results, ad-hoc to our data.
For our provenance analysis, we collected two distinct artifacts at the end of each session: the test files produced by the participants and the conversation with ChatGPT. The ChatGPT conversations are a sequence of alternating prompts (by the participant) and answers (generated by ChatGPT). % be direct and forget about web stuff
In \Cref{lst:vari}, we presented simple instances of test assertions, possibly involving multiple statements, but it is usual to interleave the statements of different test assertions and to share them. 

%Thus, to precisely analyze a significant part of our data needed to analyze tests at the level of statements, isolate the test assertions
%\andy{is ``them'' the statements, or the assertions? How I read it, it is statements (because that is the latest term you were referring to), but I think you mean assertions?}and gather the related statements, to identify and extract the semantically relevant elements of each test assertion such as the test inputs, expected values, and result (of calling the method under test).

To accurately analyze a substantial portion of our data, we needed to examine tests at the level of individual statements. This required isolating test assertions and collecting their related statements. From these, we identified and extracted the semantically relevant elements of each assertion, including the test inputs, expected values, and the result of calling the method under test.
Thus, we reached beyond purely textual approaches and leveraged syntax trees to analyze the test code~\cite{le2023hyperdiff}.
In order to infer the provenance of test assertions using a syntactic approach, we first captured the variability in tests created/generated by the participants and ChatGPT in syntactic patterns.

%To come up with the precise definition of test assertion patterns representing all the test assertions in the test files and conversations, we synthesized detailed patterns then we iteratively generalized them. 
To define test assertion patterns that represent all assertions in the test files and conversations, we first synthesized detailed patterns and then generalized them through an iterative process.
The initial synthesis involves parsing, then selecting all the code blocks (surrounded by curly braces to help isolate from ill-formed source code mixed with natural language), and synthesizing the corresponding patterns (ignoring the type identifiers and the values of literals to limit the initial number of patterns).
Then, to generalize the patterns, we iteratively apply the following four steps:
\begin{enumerate*}[label=(\roman*)]
\item we isolate the assertions (i.e., creating a new pattern for each selected assertion while removing all the other assertions in a single step). 
\item associate local variable declarations to their references syntactically through a purely syntactic approach and by comparing their names (which works on our data because in Java, a given label can only be declared once per method),
\item remove remaining local variable declarations which could not be associated to the assertion, and % the association is done recusively starting from the assertion
\item selecting the shortest list of patterns with the least number of unresolved references representing all the initial patterns.
\end{enumerate*}
% 4) then, we successively and exhaustively remove subparts from the patterns, notably blocks, statements, and argument lists, until we reach patterns with a single assertion and only the statements (e.g., local variables) it depends on.
At the end of this process, we obtain a set of abstract test patterns representing and categorizing all the assertions we collected. For example, \Cref{sublst:vari:0_initial} would be described as a local variable declaration followed by an assertion statement. While the other test assertions of \Cref{lst:vari} would be captured by the same generalized pattern described as a single assertion statement. With these patterns, we are now able to infer and describe the provenance of test assertions, controlling the patterns considered, and capturing the test elements to obtain the results of our similarity analysis. %\andy{Similarity metric is new, which also means this is a heuristic, so not exact?}
% {\color{red}We provide the detailed patterns in our data package.} %quentin to baris: there or later, here a footnote link directly pointing to the file would be good (it can replace the sentence).

Once the exhaustive set of test patterns has been obtained, we proceed with the automated inference of the origin of the tests. This process takes place in three steps, also preceded by the separation of questions and answers from conversations:
\begin{enumerate*}[label=(\roman*)]
\item parsing, 
\item selection of tests based on the previously extracted patterns, and 
\item search in the conversation in chronological order for the first match. 
\end{enumerate*}

%\andy{First time heuristic is mentioned, is it a heuristic or not, analysis might be the safer option}
The structured test patterns and provenance analysis enabled us to determine the origin of test assertions across all participants. Using this information, we can determine their dominant strategy over the implementation dimension.

\subsection{\textbf{RQ2: How do students prompt generative AI?}}

For RQ2, we extracted and analyzed the conversations between participants and ChatGPT. Similar to our approach in RQ1, we coded the prompts participants created, categorizing them based on key aspects of their interaction with ChatGPT. Specifically, we broke down each prompt into three dimensions:
    \begin{itemize}
        \item \textbf{Type:} The main purpose of a prompt; this can also be considered as its corresponding use case. Examples include asking ChatGPT to explain code, generate test cases, or clarify syntax.
        \item \textbf{Context:} The information participants provided to ChatGPT to support their request. This could include elements such as the source code file of the task, code documentation, an example test case, or a conceptual scenario for testing.
        \item \textbf{Demand:} The specific output the participant was requesting from ChatGPT, such as a test case, an explanation for a piece of code, or a refactored version of a provided code snippet.
    \end{itemize}

This breakdown 
%is needed because it 
helps us understand what participants ask ChatGPT to do and how they construct their interactions. By distinguishing between \textbf{Type}, \textbf{Context}, and \textbf{Demand}, we can identify patterns in how students leverage ChatGPT, uncover their preferred use cases, and observe whether their prompting strategies evolve over time. Additionally, this categorization enables us to assess whether students are effectively providing the necessary context to receive high-quality responses and whether their demands align with their actual needs.

\subsection{\textbf{ RQ3: What are the benefits and challenges of a generative AI-assisted test workflow for students?}}

To answer RQ3, we performed a thematic analysis~\cite{willig2017sage} of participants’ post-task interviews and triangulated these insights with our observational data. Our goal was to understand both perceived and observed experiences with ChatGPT in unit testing workflows. We analyzed the interviews and observations to identify recurring patterns related to benefits and challenges, producing distinct codes that capture how students interacted with ChatGPT throughout the tasks.

\section{Results}
\label{section:results}

This section presents the findings of our study, structured to align with the research questions. Section~\ref{GeneralQuantitativeMetrics} provides an overview of general quantitative metrics of the test suites that the participants engineered, including test coverage, assertion count, and task duration. Section~\ref{r_strategies} addresses \textbf{RQ1} by identifying the strategies students employed when integrating ChatGPT into their unit testing workflows. Section~\ref{r:prompts} answers \textbf{RQ2} through a detailed analysis of participants’ prompts, examining their types and structure. Finally, Section~\ref{post_task} focuses on \textbf{RQ3} by outlining the benefits and drawbacks of AI-assisted testing. Across all sections, we draw from multiple qualitative and quantitative data sources to provide a comprehensive view of participant behavior and reasoning.

\subsection{General Quantitative Metrics}\label{GeneralQuantitativeMetrics}

Before turning to the qualitative analysis that is needed to answer our research questions, we begin with an overview of general quantitative metrics to contextualize participant performance. Specifically, we report these metrics across three groups: the student participants in our study, the reference implementation in the Apache.utils code base, and the participants in the prior study by  Aniche et al.~\cite{AnicheTSE2022}. A detailed breakdown of these metrics, including per-participant coverage scores, test and assertion counts, and task durations, is provided in Table~\ref{tab:metrics}.

We observe that mutation score varies across the three groups. The participants of Aniche et al.'s study achieved the lowest average mutation score at 79.00\%, with a relatively high standard deviation of 12.51\%, indicating substantial variability (ranging from 62\% to 92\%). In contrast, the Apache.utils code base has the highest average mutation score at 91.00\%, with a low standard deviation of 5.77\%, showing consistent test effectiveness (ranging from 86\% to 96\%). Our participants fell in between, with an average mutation score of 88.96\% and a moderate standard deviation of 6.89, ranging from 76\% to 96\%.

We also see that the number of tests engineered varies notably.
Developers in Aniche et al.'s study created an average of 9.17 test cases per participant (ranging from 6 to 13), while the Apache.utils code base had far fewer test cases, averaging only 1.50 (although they contain many assertions). Our participants, by contrast, exhibited the highest number of tests created, averaging 12.08 test cases per participant, with a wide range from 1 to 21.

\begin{table}
\centering
\caption{Per-participant testing metrics across assignments.}
\label{tab:metrics}
\setlength{\tabcolsep}{3pt}
\begin{tabular}{llcccccccr}
\toprule
Participant & Task & Mutation & Line Cov. & Branch Cov. & Tests & Assertions & Duration  \\
\midrule
P1 & Initials & 96\% & 100\% & 100\% & 7 & 18 & 44'42''  \\
P2 & Initials & 96\% & 100\% & 100\% & 8 & 11 & 13'32''  \\
P3 & Initials & 96\% & 100\% & 100\% & 1 & 11 & 17'03''  \\
P4 & Initials & 96\% & 100\% & 100\% & 4 & 9 & 19'34''  \\
P5 & Initials & 96\% & 100\% & 95\% & 12 & 12 & 29'16''  \\
P6 & Initials & 96\% & 100\% & 100\% & 17 & 17 & 37'01''  \\
P7 & Initials & 96\% & 100\% & 100\% & 15 & 16 & 35'52''  \\
P8 & Initials & 96\% & 100\% & 100\% & 19 & 19 & 41'15''  \\
P9 & Initials & 96\% & 100\% & 100\% & 14 & 14 & 18'46''  \\
P10 & Initials & 87\% & 95\% & 95\% & 8 & 8 & 22'29''  \\
P11 & Initials & 91\% & 100\% & 100\% & 9 & 9 & 24'02''  \\
P12 & Initials & 96\% & 100\% & 100\% & 16 & 16 & 12'14''  \\
P1 & SubstringsBetween & 86\% & 100\% & 100\% & 5 & 17 & 35'27''  \\
P2 & SubstringsBetween & 86\% & 100\% & 95\% & 16 & 30 & 25'12''  \\
P3 & SubstringsBetween & 86\% & 100\% & 95\% & 1 & 14 & 16'45''  \\
P4 & SubstringsBetween & 81\% & 100\% & 90\% & 11 & 12 & 13'26'' \\
P5 & SubstringsBetween & 76\% & 96\% & 95\% & 10 & 18 & 26'15''  \\
P6 & SubstringsBetween & 76\% & 100\% & 100\% & 17 & 17 & 21'56''  \\
P7 & SubstringsBetween & 86\% & 100\% & 100\% & 17 & 21 & 42'31''  \\
P8 & SubstringsBetween & 86\% & 100\% & 100\% & 21 & 47 & 37'43''  \\
P9 & SubstringsBetween & 86\% & 100\% & 100\% & 15 & 15 & 33'05''  \\
P10 & SubstringsBetween & 81\% & 100\% & 100\% & 17 & 22 & 30'49''  \\
P11 & SubstringsBetween & 86\% & 100\% & 100\% & 13 & 14 & 20'26''  \\
P12 & SubstringsBetween & 81\% & 100\% & 95\% & 17 & 17 & 19'53''  \\
Original1 & Initials & 96\% & 100\% & 100\% & 2 & 79 & --  \\
Original1 & SubstringsBetween & 86\% & 100\% & 100\% & 1 & 24 & --  \\
Original2 & Initials & 96\% & 100\% & 100\% & 2 & 75 & --  \\
Original2 & SubstringsBetween & 86\% & 100\% & 100\% & 1 & 24 & --  \\
Aniche1 & Initials & 92\% & -- & 100\% & 11 & -- & 56'34''  \\
Aniche3 & Initials & 76\% & -- & 91\% & 6 & -- & 30'15''  \\
Aniche7 & Initials & 84\% & -- & 100\% & 12 & -- & 40'53''  \\
Aniche10 & Initials & 92\% & -- & 91\% & 13 & -- & 26'15''  \\
Aniche2 & SubstringsBetween & 68\% & -- & 93\% & 7 & -- & 27'54''  \\
Aniche13 & SubstringsBetween & 62\% & -- & 75\% & 6 & -- & 18'38''  \\
\bottomrule
\end{tabular}
\end{table}

\begin{table}[htbp]
    \centering
     \caption{Category Mapping of Ideas and Implementations }

    \begin{tabularx}{\columnwidth}{lcc}
        \toprule
        \textbf{Category Name} & \textbf{Idea} & \textbf{Implementation} \\
        \midrule
        C1 (\gptideagptimpl)          & ChatGPT   & ChatGPT   \\
        C2 (\gptideapimpl)      & ChatGPT   & Participant     \\
        C3 (\pideagptimpl)      & Participant     & ChatGPT   \\
        C4 (\pideapimpl)        & Participant     & Participant     \\
        Mental                & -               & -               \\
        Documentation         & -               & -               \\
        N\/A                   & -               & -               \\
        \bottomrule
    \end{tabularx}
    \label{tab:category_mapping}
\end{table}

\subsection{Strategies}\label{r_strategies}
Axial coding of participant activity data identified four distinct strategies, located along two dimensions.
%The axial coding on participant activity data converged on four strategies that are set on two dimensions. 
The first dimension is where the \textbf{idea} for a test case originates from, i.e, either from generative AI, or from the participant. We refer to this dimension as the semantic test case. % The possible options for this are either that the \textbf{idea} for the test case (we refer to this as semantic test case throughout the study) comes from generative AI or from the participant. 
The other dimension is \textbf{implementation}. Upon implementing a semantic test case, the code is written either by the generative AI or the participant. Therefore, the four different core strategies are:

\begin{itemize}

\item \textbf{C1} (\gptideagptimpl): This strategy refers to actions where the \textbf{idea} for the test case originates from \textbf{generative AI}, and the \textbf{implementation} is also done by Generative AI. Participants in this category rely on AI-generated test cases and AI-generated implementations. Example actions included ``asking AI to generate test cases'', ``running and fixing generated test case''.

    %GPTIDEAPIMP
    \item \textbf{C2 (\gptideapimpl)}: In this strategy, the \textbf{idea} for the test case is generated by \textbf{generative AI}, but the \textbf{implementation} is done by the \textbf{participant}. This indicates that while AI assists in generating test case ideas, participants take control of writing the actual test code. An example action is ``implementing manual test case from GPT suggestion''.
    
    %PIDEAGPTIMP
    \item \textbf{C3 (\pideagptimpl)}: This strategy involves the \textbf{participant} coming up with the test \textbf{idea}, but the \textbf{implementation} is carried out by \textbf{generative AI}. Here, participants rely on AI to generate test code based on their own conceptualization of the test case. An example action is ``asking ChatGPT to refactor test suite''. 

    %PIDEAPIMP
    \item \textbf{C4 (\pideapimpl)}: In this strategy, both the \textbf{idea} and the \textbf{implementation} of the test case are handled by the \textbf{participant}. This approach is entirely manual, with no reliance on AI for either test design or coding. Example actions included ``implementing manual test cases'', ``running and fixing implemented test cases''.
\end{itemize}

These four core strategies define the main ways in which participants interact with generative AI during the testing process, ranging from full AI dependence (C1-\cone) to fully manual testing (C4-\cfour).
Similar patterns of behavior have been observed in a different setting. In particular, Kazemitabaar et al.~\cite{kazemitabaar2023novices} conducted a study with children aged 10 to 17 who used an AI code generator while learning Python and identified four coding approaches: AI Single Prompt, AI Step by Step, Hybrid, and Manual coding. These patterns parallel the strategies we identified in this study.

In addition to the four core strategies, we identified 3 supporting categories that capture auxiliary actions performed by participants. These categories are Documentation, Mental, and N/A:

\begin{itemize}
    \item \textbf{Documentation}: This category encompasses instances where participants engage with the provided Javadoc comments in the assignment source code file to familiarize themselves with the assignment code. It involves reading and interpreting this documentation. An example action is ``reading documentation''.
    
    \item \textbf{Mental}: This refers to moments where participants are actively thinking about the code, analyzing its structure, or reading through the code to develop a mental model for understanding. This cognitive process helps participants break down and internalize the logic of the code. Example actions are ``looking at method code body and precondition'' and ``considering completeness of test suite''.

    \item \textbf{N/A (Not Applicable)}: These activities include actions taken by participants that are not directly related to task advancement but are instead aimed at resolving personal difficulties, such as asking for clarification on a method or posing a question about Java syntax. Since these actions do not influence the participant’s overall strategy and are specific to their individual understanding, we exclude them from the strategy discussion. An example action is ``asking ChatGPT about Java syntax''. %In %Figure \ref{fig:flows}, they are represented in white.

\end{itemize}

\begin{figure}[!t]
    \centering
    \includegraphics[width=\textwidth, trim=0.0cm 1cm 0.0cm 4cm, clip]{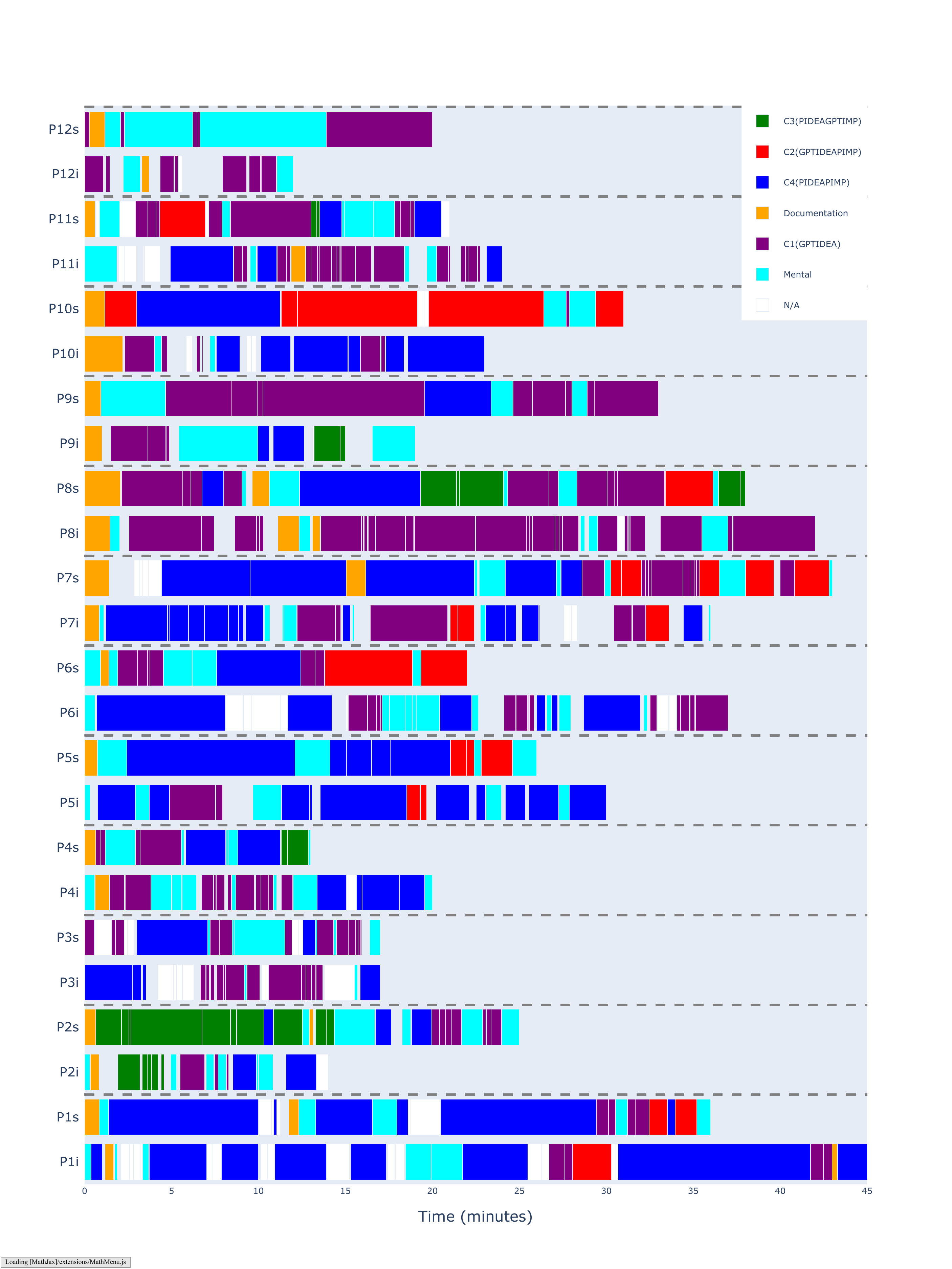}
    %\vspace{-10pt}
    \caption{Participant Approaches to Testing}
    \label{fig:flows}
\end{figure}

Figure~\ref{fig:flows} provides a detailed breakdown of participants’ actions during the experiment, organized per assignment (e.g., P3i refers to participant~3’s task involving the \textit{initials} method, while P3s refers to the task involving the \textit{substringsBetween} method). Now that we have explained all strategies (C1–C4) and auxiliary actions (documentation, mental, N/A), we turn to identifying the dominant component of a participant’s strategy for creating unit tests. An intuitive approach is to observe the most frequently occurring strategic action, either in terms of the number of actions or the total duration spent on that action throughout the task. However, a major caveat with this type of categorization is the disparity in terms of produced test cases between strategies that involve automated test creation (C1-\cone) and those that involve fully manual test creation (C4-\cfour).

For example, if a participant performs both C1 and C4 actions twice, the resulting test suite might still be primarily composed of generated tests. This suggests that automated actions can have a larger impact on the test suite compared to manual ones. To account for this, we first assess the direct contribution of ChatGPT to the participants’ final test suites by figuring out which assertions present in the test suite were suggested by ChatGPT, allowing us to evaluate the actual influence of generated tests. This analysis provides a measure of the participant’s reliance on ChatGPT.

We use this ChatGPT reliance to distinguish between strategies dominated by automated actions (C1 and C3) and those driven more by manual efforts (C2 and C4). Following this initial separation, we further differentiate these strategy groups by examining the frequency of performed actions. This process enables us to assign a dominant strategy to each participant, or in cases where no single strategy dominates, to classify them as employing a mixed strategy.

To infer the origin of a test as suggested by ChatGPT, we relied on the structure of the conversations and on the patterns we extracted (see Section~\ref{section:assertionmatching}). For each test file and its corresponding ChatGPT conversation, we extracted the test assertions along with their related statements. Each assertion in the test file was then iteratively compared to each assertion in the conversation in chronological order. 
If we find a match in an answer generated by ChatGPT, then we consider the assertion as generated. If we do not find a match, we consider the assertion to be manually implemented.
%Once a match was found, if it was an answer generated by ChatGPT, then the assertion was considered as generated, otherwise if it was a prompt the assertion would have to be considered as manual.% (we cannot distinguish easily anymore between the assistant generating a novel test or simply repeating the previouly provided test). 

% it is necessary to keep the order of QA in convs to find who first introduce the test
% \quentin{\textbf{[good to reexplain, TODO reexplain while getting an overview of participant\_summary and using a concrete example/walkthough ]} Once the exhaustive set of test patterns has been obtained, we proceed with the automated inference of the origin of the tests. This process takes place in three steps, also preceded by the separation of questions and answers from conversations : 1) parsing, 2) selection of tests based on the previously extracted patterns, 3) search in the conversation in chronological order for the first match. 
% }
% \quentin{\textbf{[Do not need that much details anymore, it will be modified to describe table with assertion provenance numbers]}

\Cref{tab:participant_summary} presents the results of our analysis and enables us to compute the overall percentage of assertions that we inferred to originate from the ChatGPT conversations.
This table also provides the number of test assertions present in the test files.
Additionally, based on our analysis of test patterns, we also provide the breakdown of a subset of patterns representing the examples in \Cref{lst:vari} where the assertion requires from zero to any number of local variable declaration and we only compare the inputs (ignoring the expected value). %quentin to baris: do we explain that in the methodology or you consider it a detail ?

We omitted the breakdown of more complex and less frequently occurring test patterns, such as ones requiring for loops or assignment statements. In particular, we did not analyze $\sim$10\% of the test assertions in the test files of our participants. Participant 1 (P1) represents 
%On the other hand, this breakdown represents 90\% (360 out of 404) of the test assertions in the test files of our participants.
P1 represents 
7.6\% (31 of 404) of non-detailed patterns, which is due to P1 mostly using parametric testing, thus separating the assertions from their inputs across different methods and making them difficult to analyze. The remaining non-detailed assertions only represent 0.3\% (13 of 404) of the total.

We present the detailed provenance ratios as fractions of the number of assertions originating from ChatGPT over the total number of assertions in the test file for the given pattern.
The detailed provenance ratios are presented from instances with no local variable declaration (a single assertion statement) to five, where a match only occurs between instances with the same number of local variables.
On the first column of the breakdown, the single statement test assertions represent 46\% of assertions in test files.
The last column of the breakdown presents the result of combining all the previous patterns, such that additional matches in this column represent instances where a test assertion was refactored to use a different number of local variables. The details and intermediate results can be found in our data package \cite{datapackage}.%quentin to baris: footnote or only in the methodology or elsewhere ? % quentin: you choose, but this part references the provenance subcommand which uses the patterns in pattern/*.scm

% We highlight the patterns with over X occurrences in the final test files, accounting for XX\% of assertions, while the remaining results are aggregated in a single fraction.
% This table also presents the strategy of each assignment that we inferred from our results.

% To take into account the slight variations in the tests and detail the different processes of test copy-pasting by the participants, we distinguish 4 different heuristics for matching the tests: 1) the block containing the assertion match exactly, 2) the assertion statement contains all these arguments (no access to local variables) and match exactly, 2a) the expected value changes \andy{What do you mean by expected value changes?}, 2b) the entry of the tested function changes, 3) the assertion statement uses local variables, all the statements match exactly, 3a) the expected value changes, 3b) the entry of the tested function changes. 
%  \andy{what's the difference between heuristic 2 and 3?}
% Our results are presented in \Cref{tab:participant_summary} and detail the number of tests according to the inference heuristic. 
% We observe that it cross-validates with all our video analysis of the programming sessions.
% The detailed table is provided in our reproduction package.

\begin{table}
\newcommand{\assrt}[3]{%
\ifboolexpr{
    test {\ifnumequal{#1}{0}} 
    and test {\ifnumequal{#2}{0}}
  }
  {}%
  {\sfrac{#1}{#2}}%
% \sfrac{#1}{#2}\textsuperscript{#3}%
}
\centering
\caption{Participant Strategy and Assertion Percentages}
\label{tab:participant_summary}
\begin{tabular}{c@{\hspace{2mm}}cc@{\hspace{1mm}}c@{\hspace{1mm}}c@{\hspace{1mm}}c@{\hspace{1mm}}c@{\hspace{1mm}}c@{\hspace{1mm}}c@{\hspace{4mm}}ccc}
\toprule
\multirow{2}{*}{Assignment} &  \multirow{2}{*}{Strategy} & \multicolumn{7}{c}{Assertions per \#local var.} & \multicolumn{2}{c}{Assertion} \\
                            &                            & 0 & 1 & 2 & 3 & 4 & 5 &  $\le$5                            &      count    &   \% gen      \\
\midrule                                                    

    P12s &       C1 (\cone)   & \assrt{17}{17}{0} & \assrt{0}{0}{1}   & \assrt{0}{0}{2} & \assrt{0}{0}{3}  &  \assrt{0}{0}{4}  & \assrt{0}{0}{5} & \assrt{17}{17}{n} & 17 & 100\%  \\% &  58.8\% \\        
    P12i &       C1 (\cone)   & \assrt{16}{16}{0} & \assrt{0}{0}{1}   & \assrt{0}{0}{2} & \assrt{0}{0}{3}  &  \assrt{0}{0}{4}  & \assrt{0}{0}{5} & \assrt{16}{16}{n} & 16 & 100\%  \\% &  68.8\% \\       
    P11s &       C1 (\cone)   & \assrt{5}{10}{0}  & \assrt{4}{4}{1}   & \assrt{0}{0}{2} & \assrt{0}{0}{3}  &  \assrt{0}{0}{4}  & \assrt{0}{0}{5} & \assrt{9}{14}{n}  & 14 & 64.3\% \\% &  42.9\% \\        
    P11i &       C1 (\cone)   & \assrt{1}{1}{0}   & \assrt{1}{2}{1}   & \assrt{2}{4}{2} & \assrt{2}{2}{3}  &  \assrt{0}{0}{4}  & \assrt{0}{0}{5} & \assrt{6}{9}{n}   & 9  & 66.7\% \\% &  66.7\% \\        
    P10s &       C2 (\ctwo)   & \assrt{0}{0}{0}   & \assrt{0}{0}{1}   & \assrt{0}{0}{2} & \assrt{0}{11}{3} &  \assrt{0}{11}{4} & \assrt{0}{0}{5} & \assrt{0}{22}{n}  & 22 & 0.\%   \\% &   0.\%  \\   
    P10i &      C4 (\cfour)   & \assrt{0}{0}{0}   & \assrt{0}{0}{1}   & \assrt{1}{8}{2} & \assrt{0}{0}{3}  &  \assrt{0}{0}{4}  & \assrt{0}{0}{5} & \assrt{1}{8}{n}   & 8  & 12.5\% \\% &   0.\%  \\     
     P9s &       C1 (\cone)   & \assrt{13}{15}{0} & \assrt{0}{0}{1}   & \assrt{0}{0}{2} & \assrt{0}{0}{3}  &  \assrt{0}{0}{4}  & \assrt{0}{0}{5} & \assrt{13}{15}{n} & 15 & 86.7\% \\% &  53.3\% \\        
     P9i &            Mixed   & \assrt{12}{14}{0} & \assrt{0}{0}{1}   & \assrt{0}{0}{2} & \assrt{0}{0}{3}  &  \assrt{0}{0}{4}  & \assrt{0}{0}{5} & \assrt{12}{14}{n} & 14 & 85.7\% \\% &  78.6\% \\        
     P8s &       C1 (\cone)   & \assrt{0}{0}{0}   & \assrt{24}{47}{1} & \assrt{0}{0}{2} & \assrt{0}{0}{3}  &  \assrt{0}{0}{4}  & \assrt{0}{0}{5} & \assrt{24}{47}{n} & 47 & 51.1\% \\% &  40.4\% \\        
     P8i &       C1 (\cone)   & \assrt{4}{10}{0}  & \assrt{0}{0}{1}   & \assrt{6}{7}{2} & \assrt{2}{2}{3}  &  \assrt{0}{0}{4}  & \assrt{0}{0}{5} & \assrt{12}{19}{n} & 19 & 63.2\% \\% &  47.4\% \\        
     P7s &      C4 (\cfour)   & \assrt{0}{6}{0}   & \assrt{0}{10}{1}  & \assrt{0}{0}{2} & \assrt{0}{0}{3}  &  \assrt{0}{1}{4}  & \assrt{0}{4}{5} & \assrt{1}{21}{n}  & 21 & 4.8\%  \\% &   0.\%  \\    
     P7i &      C4 (\cfour)   & \assrt{0}{15}{0}  & \assrt{0}{1}{1}   & \assrt{0}{0}{2} & \assrt{0}{0}{3}  &  \assrt{0}{0}{4}  & \assrt{0}{0}{5} & \assrt{0}{16}{n}  & 16 & 0.\%   \\% &   0.\%  \\   
     P6s &            Mixed   & \assrt{9}{17}{0}  & \assrt{0}{0}{1}   & \assrt{0}{0}{2} & \assrt{0}{0}{3}  &  \assrt{0}{0}{4}  & \assrt{0}{0}{5} & \assrt{9}{17}{n}  & 17 & 52.9\% \\% &  47.1\% \\        
     P6i &            Mixed   & \assrt{9}{14}{0}  & \assrt{1}{1}{1}   & \assrt{0}{0}{2} & \assrt{0}{0}{3}  &  \assrt{0}{0}{4}  & \assrt{0}{0}{5} & \assrt{10}{15}{n} & 17 & 33.3\% \\% & 33.3\%  \\      
     P5s &      C4 (\cfour)   & \assrt{0}{7}{0}   & \assrt{0}{1}{1}   & \assrt{0}{6}{2} & \assrt{0}{0}{3}  &  \assrt{0}{0}{4}  & \assrt{0}{0}{5} & \assrt{0}{14}{n}  & 18 & 0.\%   \\% &   0.\%  \\    
     P5i &      C4 (\cfour)   & \assrt{0}{11}{0}  & \assrt{0}{0}{1}   & \assrt{0}{0}{2} & \assrt{0}{0}{3}  &  \assrt{0}{0}{4}  & \assrt{0}{0}{5} & \assrt{0}{11}{n}  & 12 & 0.\%   \\% &   0.\%  \\    
     P4s &            Mixed   & \assrt{0}{0}{0}   & \assrt{7}{12}{1}  & \assrt{0}{0}{2} & \assrt{0}{0}{3}  &  \assrt{0}{0}{4}  & \assrt{0}{0}{5} & \assrt{7}{12}{n}  & 12 & 58.3\% \\% &  100\%  \\       
     P4i &       C1 (\cone)   & \assrt{0}{1}{0}   & \assrt{7}{8}{1}   & \assrt{0}{0}{2} & \assrt{0}{0}{3}  &  \assrt{0}{0}{4}  & \assrt{0}{0}{5} & \assrt{7}{9}{n}   & 9  & 77.8\% \\% &  77.8\% \\        
     P3s &       C1 (\cone)   & \assrt{9}{10}{0}  & \assrt{0}{0}{1}   & \assrt{0}{0}{2} & \assrt{0}{0}{3}  &  \assrt{0}{0}{4}  & \assrt{0}{0}{5} & \assrt{9}{10}{n}  & 14 & 64.3\% \\% &  66.7\% \\        
     P3i &       C1 (\cone)   & \assrt{8}{10}{0}  & \assrt{0}{0}{1}   & \assrt{0}{0}{2} & \assrt{0}{0}{3}  &  \assrt{0}{0}{4}  & \assrt{0}{0}{5} & \assrt{8}{10}{n}  & 11 & 72.7\% \\% &  45.5\% \\        
     P2s &     C3 (\cthree)   & \assrt{5}{7}{0}   & \assrt{13}{20}{1} & \assrt{0}{0}{2} & \assrt{0}{1}{3}  &  \assrt{1}{1}{4}  & \assrt{0}{0}{5} & \assrt{19}{29}{n} & 30 & 63.3\% \\% &  26.7\% \\        
     P2i &            Mixed   & \assrt{0}{0}{0}   & \assrt{1}{3}{1}   & \assrt{5}{8}{2} & \assrt{0}{0}{3}  &  \assrt{0}{0}{4}  & \assrt{0}{0}{5} & \assrt{6}{11}{n}  & 11 & 54.5\% \\% &  36.4\% \\        
     P1s &      C4 (\cfour)   & \assrt{0}{1}{0}   & \assrt{0}{0}{1}   & \assrt{0}{0}{2} & \assrt{0}{0}{3}  &  \assrt{0}{0}{4}  & \assrt{0}{0}{5} & \assrt{0}{1}{n}   & 17 & 0.\%   \\% &   0.\%  \\   
     P1i &      C4 (\cfour)   & \assrt{0}{2}{0}   & \assrt{0}{1}{1}   & \assrt{0}{0}{2} & \assrt{0}{0}{3}  &  \assrt{0}{0}{4}  & \assrt{0}{0}{5} & \assrt{0}{3}{n}   & 18 & 0.\%   \\% &   0.\%  \\  
\bottomrule
\end{tabular}
\end{table}

\paragraph{C1 (\gptideagptimpl) dominant:}
Participants \textbf{P3, P8, P11, and P12} consistently relied on ChatGPT for both assignments. \textbf{P4}, on the other hand, initially employed a \textbf{C1} dominant strategy but transitioned to a mixed strategy for the second assignment. \textbf{P9} did the opposite and moved from a mix to a \textbf{C1} dominant approach. As a result, we observed that about half of the participants primarily utilized \textbf{C1} in at least one of their assignments. 

\paragraph{C2 (\gptideapimpl) dominant:}
This is the third most common strategy, following \textbf{C1} and \textbf{C4}. It was the dominant approach in the second assignment for \textbf{P10}. Additionally, it appeared in the assignments of \textbf{P1}, \textbf{P5}, and \textbf{P7}, though it was secondary to \textbf{C4}.

\paragraph{C3 (\pideagptimpl) dominant:}
This strategy was the least frequently used, emerging as dominant only in the second assignment for \textbf{P2}. It was also observed in the second assignment of \textbf{P8} and \textbf{P4}, as well as the first assignment of \textbf{P9} and \textbf{P2}.

\paragraph{C4 (\pideapimpl) dominant:} This was the second most common strategy. Traditional programming was the preferred approach for \textbf{P1, P5, and P7} in both assignments, as well as in the first assignment for \textbf{P10}. \textbf{P12} was the only participant who did not engage in any \textbf{C4}-type actions.

\paragraph{Mixed:}
Participants \textbf{P2, P4, P6, and P9} exhibited mixed strategies in their first assignments, where no single approach was clearly dominant. The most frequently observed combination was the \textbf{C1-C4} pair.
\paragraph{Documentation:}
The documentation category is primarily observed at the beginning of assignments for most participants. However, it is absent for P3 and does not appear in the first assignment for P5, though it is present in the second. In general, documentation-related actions occur early in the assignment and are almost always within the first half. Most participants interact with the Javadoc only once, with a maximum of three interactions. This suggests that participants prioritize reviewing documentation at the start of their tasks, aligning with findings from Aniche et al.~\cite{aniche2019pragmatic}.
\paragraph{Mental:}
The mental category is present for all participants. The proportion of time spent on this activity, relative to the overall assignment duration, varies between participants, ranging from 8\% to 40\%. Regardless of the dominant strategy for creating their test suites, we observed that participants attempted to build a mental model of code.

\paragraph{Summary for RQ1:}
We observed four core strategies (C1-\cone to C4-\cfour) characterizing how participants engaged with ChatGPT during unit test creation, defined along two dimensions: the origin of the test case idea and the implementation source. These strategies range from full reliance on ChatGPT (C1-\cone) to fully manual workflows (C4-\cfour). To determine each participant’s dominant approach, we combined strategy frequency with automated provenance tracing of test assertions. This allowed us to assess not just action frequency but also the concrete contribution of ChatGPT to the final test suites. In addition to these strategies, we identified three auxiliary activity types: documentation, mental modeling, and non-task-related actions, that contextualize participant behavior beyond direct test creation. Together, these findings offer a comprehensive account of participant strategies and levels of ChatGPT reliance.

\subsection{Prompts}\label{r:prompts}

\begin{figure}[!t]
    \centering
    %\includesvg[width=\linewidth]{pictures/sankeyTCG.svg}
    \includegraphics[width=\textwidth]{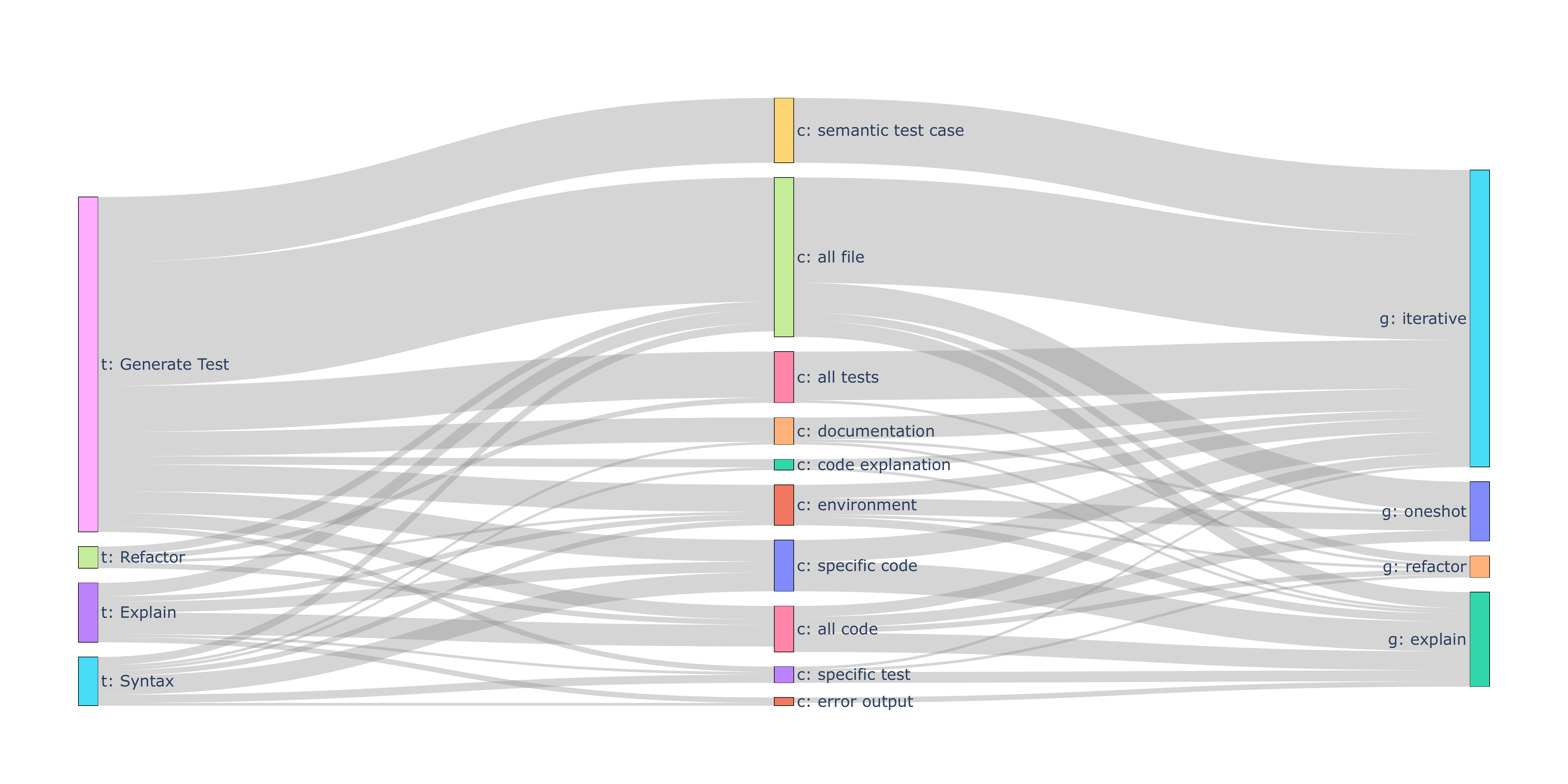}
    %\vspace{-10pt}
    \caption{Breakdown of participants' prompts by type (t), provided context (c), and grouped demand (g), where g reflects consolidated output requests}
    \label{fig:sankey}
\end{figure}

%For the analysis we have broken down prompts into 3 segments that are "type", "context" and "demand". 
%Prompt types inform as regarding the general use cases. Context allows us to see what participants choose to provide to GPT. Demand demonstrates the reasons they had for interacting with GPT. We have further categorized test generation strategies of participants by grouping up their approaches into 2 which are oneshot or iterative generation of tests. These reflect the behaviour of participants for test generation where oneshot strategy opts for generating a test suit following with tasks to refine it while iterative build the test suit over time with the generation of single or multiple tests with ongoing refinement until a satisfying test suite is reached. 

We investigated a total of 111 prompts submitted across all assignments. On average, participants wrote 9.25 prompts per assignment. The number of prompts per participant ranged from a minimum of 3 to a maximum of 17, reflecting variation in how students used ChatGPT across tasks. This prompt volume provides a rich basis for investigating how students engaged with ChatGPT on what types of requests they made, how much context they provided, and the overall structure of their interactions.

To analyze the participant prompts during the assignments, we have divided each prompt into three components: type, context, and demand. These are visualized in Figure~\ref{fig:sankey}. These components help us understand how students interact with ChatGPT during unit testing tasks, including why they interact with the tool (type), what they ask for (demand), and what information they provide to it (context). 

\subsubsection{Prompt Types}

The prompt types reflect the general use cases. The dominant type was \textbf{test generation} (72 out of 111), indicating that the primary use case for ChatGPT was to assist with the creation of unit tests. Other uses included \textbf{explaining} code or test behavior (17), dealing with \textbf{syntax-level} concerns (15), and performing \textbf{refactorings} on existing test code (7). This breakdown reveals that ChatGPT’s primary role was in test generation, but that it also served as a flexible assistant throughout the assignments.

\subsubsection{Prompt Demand and Test Generation Strategies}

\begin{table}[!t]
\centering
\caption{Prompt Demand Group Breakdown by Participant-Assignment Pair }
\label{tab:demandgroupbreakdown}
\begin{tabular}{lcccc}
\toprule
 Participant   &   Explain &   Oneshot &   Iterative &   Refactor \\
\midrule
 P1i            &         8 &          0 &           1 &          0 \\
 P2i            &         0 &          0 &           2 &          0 \\
 P3i            &         3 &          1 &           3 &          2 \\
 P4i            &         2 &          1 &           1 &          1 \\
 P5i            &         0 &          1 &           1 &          0 \\
 P6i            &         2 &          1 &           0 &          0 \\
 P7i            &         3 &          0 &           1 &          0 \\
 P8i            &         1 &          0 &           5 &          0 \\
 P9i            &         0 &          1 &           0 &          1 \\
 P10i           &         1 &          1 &           2 &          0 \\
 P11i           &         3 &          0 &           8 &          0 \\
 P12i           &         0 &          1 &           1 &          0 \\
 P1s            &         1 &          0 &           2 &          0 \\
 P2s            &         0 &          0 &           7 &          0 \\
 P3s            &         2 &          1 &           3 &          2 \\
 P4s            &         0 &          1 &           0 &          1 \\
 P5s            &         0 &          1 &           0 &          0 \\
 P6s            &         3 &          2 &           3 &          0 \\
 P7s            &         1 &          0 &           6 &          0 \\
 P8s            &         0 &          1 &           4 &          0 \\
 P9s            &         0 &          1 &           2 &          0 \\
 P10s           &         1 &          0 &           2 &          0 \\
 P11s           &         2 &          1 &           3 &          0 \\
 P12s           &         0 &          1 &           1 &          0 \\
\bottomrule
\end{tabular}
\end{table}

Prompt demand captures the participants’ goals when interacting with ChatGPT. The three dominant categories were:

\begin{itemize}
    \item \textbf{Explain} (33): These prompts sought clarification about various aspects of the code or testing process, including syntax, behavior, and results.
    \begin{quote}
        Within this category, the most common explanations requested were related to \textbf{code or method behavior} (14 instances) and \textbf{syntax} (13 instances). Less frequent but notable were prompts asking for explanations of \textbf{code output}~(2) or \textbf{test results}~(2), indicating a need to interpret what the system produced in response to prior inputs.
    \end{quote}

    \item \textbf{Iterative test generation} (71): These involved generating individual test cases or small groups of tests in a step-by-step fashion.
    \begin{quote}
        Within the category of iterative test generation, we identified several specific types of demands that reflect how students used ChatGPT to build and refine their test suites progressively. The most frequent demands were to \textbf{generate specific test case} (28 instances) and \textbf{expand an existing test suite} (16 instances), highlighting a common strategy of incrementally improving coverage based on previously existing tests, which were either generated, manually written or both.

        Participants also requested generation of \textbf{edge cases} (9). Four prompts involved asking for \textbf{semantic test cases}, indicating that participants were also providing descriptions of tests that they wanted to generate.

        Other iterative demands included requests for a \textbf{specific test format} (5) and a \textbf{specific number of tests to generate} (6). Though less frequent, additional demands included generation of \textbf{example test inputs} (2) and explicit \textbf{non-edge case} scenarios (1).

        These patterns indicate that students using iterative prompting often approached ChatGPT as a collaborator, using it to create the test suite in a structured, layered manner rather than through oneshot generation. This led to participants using more prompts as demonstrated in Table \ref{tab:demandgroupbreakdown}. 
    \end{quote}

    \item \textbf{Oneshot test generation} (16): These prompts asked ChatGPT to generate a full test suite in a single prompt.
    \begin{quote}
        Oneshot prompts are fewer in number, as demonstrated in Table \ref{tab:demandgroupbreakdown}. This is expected because each oneshot prompt delivers an entire test suite, whereas iterative prompting divides the process into multiple, smaller requests. Thus, the lower frequency of oneshot prompts should not be interpreted as a lack of interest, but rather as a reflection of a different engagement style.
    \end{quote}
\end{itemize}

\subsubsection{Prompt Context}

The prompt context refers to the kind of information participants included with their prompts. Out of a total of 110 prompts, participants included a wide variety of contextual elements to help the model generate more relevant and accurate outputs.

The most common form of context was \textbf{all file} inclusion (57 instances), where participants supplied the entire source file with the JavaDoc documentation, to give ChatGPT a comprehensive understanding of the codebase. This was followed by \textbf{semantic test case} descriptions (24), indicating that some participants preferred to describe what the test to be generated should do, rather than relying on providing source code.

Participants also frequently provided \textbf{specific code} snippets (19) or included \textbf{environment} details (15) like the programming language or unit test framework they used (Java and JUnit).
In 19 prompts, participants also included \textbf{all tests} from their current test suites at the time of prompting.
This was typically done toward the end of the interaction, either to help ChatGPT iteratively refine the test suite or to assess its completeness. In 9 prompts, participants provided a \textbf{specific test} or a subset of tests from their suite. These were most often provided as examples to guide the generation of similar test cases or to demonstrate a preferred format or template. 

Less frequently, students included \textbf{documentation} (10), \textbf{code explanations} (4), \textbf{error outputs} (3), and \textbf{specific test cases} (6). These more focused forms of context show that some students were attempting to fine-tune ChatGPT’s understanding in order to resolve specific issues such as debugging, troubleshooting, or obtaining detailed clarifications.

Overall, this diversity in context types reflects flexible prompting strategies. While some students opted for completeness by providing full files, for example, providing the entire task file, which includes code under test, auxiliary methods, and Javadoc, others took a more targeted approach, sharing code snippets or specific semantic test cases, guiding ChatGPT with precise information. Semantic test cases, in particular, indicate that students often engaged with ChatGPT at a conceptual level, describing what should be tested rather than only providing code under test.

%Participants often provided targeted input to guide GPT’s responses, rather than relying on generic prompts. Semantic test cases, in particular, indicate that students often engaged with ChatGPT at a conceptual level—describing what should be tested rather than only providing code under test. 

\subsubsection{Oneshot vs. Iterative Prompting Strategies}

To determine whether distinct patterns exist in how participants interact with ChatGPT, we constructed state diagrams for each participant-assignment pair. These diagrams are available in our data package~\cite{datapackage}. These diagrams visualize the sequence and structure of prompts, enabling us to identify systematic differences in engagement styles. Through this analysis, we inductively derive two main prompting strategies: \textbf{oneshot}, where participants requested a complete test suite in a single prompt, and \textbf{iterative}, where test generation was approached through a sequence of smaller, focused prompts. We further categorized participants based on their test generation strategy: \textbf{oneshot} vs. \textbf{iterative}. A breakdown of this categorization is available in Table~\ref{tab:promptingStrategy}.

\begin{table}[h]
\centering
\caption{Participant-Assignment pairs by prompting strategy.}
\begin{tabular}{cc}
\toprule
\textbf{Iterative Strategy} & \textbf{Oneshot Strategy} \\
\midrule
\texttt{P1i, P1s} & \texttt{P3i, P3s} \\
\texttt{P2i, P2s} & \texttt{P4i, P4s} \\
\texttt{P7i, P7s} & \texttt{P5i, P5s} \\
\texttt{P8i}      & \texttt{P6i, P6s} \\
\texttt{P10s}     & \texttt{P8s} \\
\texttt{P11i}     & \texttt{P9i, P9s} \\
                 & \texttt{P10i} \\
                 & \texttt{P11s} \\
                 & \texttt{P12i, P12s} \\
\bottomrule
\end{tabular}
\label{tab:promptingStrategy}
\end{table}

%This division shows that half of the participants (6 out of 12) leaned toward a oneshot test generation approach, treating GPT as a tool to deliver a comprehensive solution. However, a notable minority (3 participants) preferred iterative prompting, indicating a more interactive and gradual process where GPT’s outputs were actively refined and validated over time. In particular, 3 other participants (P8, P10, and P11) demonstrated mixed behavior as they used different strategies in different assignments. 
This division reveals that half of the participants (6 out of 12) followed a oneshot test generation approach, treating ChatGPT as a tool for producing complete solutions. In contrast, 3 participants used iterative prompting, engaging in a more interactive and gradual process where ChatGPT’s outputs were actively refined and validated over time. Additionally, the remaining 3 participants (P8, P10, and P11) exhibited mixed behavior, using different strategies across assignments.

\paragraph{Summary for RQ2:}

Students engaged with ChatGPT in multifaceted ways, using it not only for generating test cases but also for explaining logic, clarifying syntax, and refining outputs. Most prompts were rooted in meaningful contexts, and participants exhibited a variety of prompting styles; some preferred quick, oneshot generation, while others engaged in more iterative and collaborative workflows. While the short nature of the assignments limits our ability to track prompt evolution over time, the data suggests that students were mostly consistent in how they used ChatGPT during these assignments.

\begin{table}[h]
\footnotesize % Smaller font for tighter fit
\caption{Summary of Perceived Drawbacks and Benefits}
\setlength{\tabcolsep}{4pt} % Tighter column padding
\centering
\newcolumntype{L}[1]{>{\raggedright\arraybackslash}p{#1}}
\newcolumntype{C}[1]{>{\centering\arraybackslash}p{#1}}

% Adjust max bar width for narrower column
\newcommand{\maxfreq}{13}
\newcommand{\maxbarwidth}{3.5cm}

\newcommand{\frequencybarinline}[1]{%
    \begin{tikzpicture}[baseline]
        \fill[blue!50] (0,0) rectangle (#1*\maxbarwidth/\maxfreq,0.25);
        \draw[black] (0,0) rectangle (\maxbarwidth,0.25);
    \end{tikzpicture}%
}

\begin{tabular}{@{}L{3.5cm}C{4.8cm}@{}} % Total width ~8cm
\toprule
\textbf{Category} & \textbf{Count} \\
\midrule

\textbf{Benefits} & \\
\textit{Cognitive Load} & 10 \frequencybarinline{10} \\
\textit{ChatGPT as Supervisor} & 6 \frequencybarinline{6} \\
\textit{Participant as Supervisor} & 8 \frequencybarinline{8} \\
\textit{Saves Time} & 13 \frequencybarinline{13} \\
\textit{Test Generation} & 6 \frequencybarinline{6} \\
\textit{ChatGPT Expands Search on New Subject} & 1 \frequencybarinline{1} \\
\textit{Uses ChatGPT for Knowledge Questions} & 1 \frequencybarinline{1} \\
\textit{ChatGPT Explains Methods Well} & 1 \frequencybarinline{1} \\
\textit{Uses ChatGPT for Syntax} & 3 \frequencybarinline{3} \\
\midrule
\textbf{Drawbacks} & \\
\textit{Lack of Ownership} & 6 \frequencybarinline{6} \\
\textit{Lack of Quality} & 13 \frequencybarinline{13} \\
\textit{Lack of Trust} & 9 \frequencybarinline{9} \\
\textit{Loss of Skill} & 1 \frequencybarinline{1} \\
\textit{Negative Belief} & 1 \frequencybarinline{1} \\
\textit{Prevent Learning} & 1 \frequencybarinline{1} \\
\bottomrule
\end{tabular}

\label{tab:drawbacks_and_benefits}
\end{table}

\subsection{Post Task Interviews}\label{post_task}

To address \textbf{RQ3} — \textit{What are the benefits and challenges of a generative AI-assisted test workflow for students?}, we conducted a thematic analysis of participants’ post-task interviews, supported by observational data. This analysis revealed recurring patterns in how students interacted with ChatGPT during the testing tasks.

Several benefits and drawbacks emerged from this investigation, particularly regarding productivity, support, ownership, and the nature of collaboration between participants and ChatGPT. These are summarized in Table~\ref{tab:drawbacks_and_benefits}.

\subsubsection{Benefits}\label{r_benefits}

The following themes capture the key ways in which students found ChatGPT beneficial during their unit testing tasks. These benefits range from gains in efficiency and reduced mental effort to more subjective forms of support and collaborative interaction.

\paragraph{Saves Time}

%\subsubsection{Saves Time}
\textbf{Time-saving} was the most frequently mentioned benefit (13 mentions). Participants found that ChatGPT accelerated their workflow by quickly generating boilerplate or initial versions of test cases, allowing them to iterate faster or redirect time to more critical thinking tasks. This aligns with the general perception of AI as a productivity enhancer.

\paragraph{Reduced Cognitive Load}
Participants frequently reported that ChatGPT helped reduce their \textbf{cognitive load} (10 mentions). By offloading some repetitive aspects of test creation, students could focus more on understanding the application logic or coming up with more test cases after covering the basic functionality of the code under test quickly. This can make unit testing less mentally taxing if generative AI use does not detriment personal skill development, especially for less experienced programmers.

This aligns with findings by Camilleri et al.~\cite{camilleri2022investigating}, who observed that test case design is one of the testing tasks associated with significant cognitive workload, especially among less experienced software testers. Using the NASA-TLX instrument, this study showed elevated levels of mental demand, effort, and frustration in such tasks, confirming that reducing the load associated with test case creation could offer practical benefits.

\paragraph{Supervision Dynamics – Human and AI}
We identified two distinct yet complementary supervisory dynamics that capture the participants’ overall approach to collaborating with AI for unit testing. Some participants perceived \textbf{ChatGPT as a supervisor} (6 mentions), guiding them through the unit testing process, suggesting test logic, or explaining unfamiliar concepts. Conversely, others viewed themselves as the supervisor of the AI (\textbf{Participant as Supervisor}, 8 mentions), feeling empowered to critique, refine, or selectively use ChatGPT’s suggestions. This reflects a more interactive and agent-driven relationship with the tool.

\paragraph{Test Generation Assistance}
ChatGPT’s ability to assist with \textbf{test generation} (6 mentions) was highlighted as a key benefit. While concerns about test quality persist, many students appreciated having a starting point from which they could build or improve test cases. The tool serves more as a productivity aid than a complete solution, as the generated tests frequently contain errors in setting up the expected values for assertions. 

\paragraph{Targeted Use Cases}
Some participants reported less frequent use cases of ChatGPT:
	It was used much like a search engine to \textbf{expand their search} for new information (1 mention), to answer \textbf{knowledge-based questions} (1 mention) such as understanding testing concepts. Participants also used it to explain methods or APIs clearly (\textbf{ChatGPT explains methods well}, 1 mention) and to address \textbf{syntax-level} challenges (3 mentions), helping overcome low-level barriers that might otherwise stall progress.

These niche benefits demonstrate how ChatGPT can flexibly integrate into different parts of a student’s workflow, addressing both high-level reasoning and low-level mechanics.

\subsection{Drawbacks}\label{r_drawbacks}

While AI-assisted tools like ChatGPT offer several perceived benefits in the testing workflow, participants also reported a range of concerns and limitations. These drawbacks provide critical insights into the potential risks of over-reliance on automation and underscore the need for careful integration of AI into learning environments. Similar concerns were also emphasized by Choudhuri et al.~\cite{choudhuri2024insights} 

\paragraph{Lack of Ownership}
Several participants (6 mentions) expressed a \textbf{lack of ownership} over the test cases generated by ChatGPT. For the tests that were not written manually, students felt less accountable for their correctness and were more inclined to accept the output without critical evaluation. This detachment can reduce opportunities for active learning and reflection, both of which are essential for developing a strong understanding of unit testing principles.
Prior work has similarly noted that perceived ownership tends to increase when learners are more actively involved in generating solutions or when they are reminded of their involvement through prompting and refinement~\cite{wasi2024llms}.

\paragraph{Lack of Quality}
The most frequently cited drawback (13 mentions) was the \textbf{lack of quality} in AI-generated test cases. Participants noted that these tests often:
\begin{itemize}
\item Miss edge cases or complex logic.
\item Do not align well with functional requirements.
\item Are unpolished and contain duplicates.
\end{itemize}
This suggests that while ChatGPT can generate syntactically correct test cases, human oversight is essential to ensure relevance and depth. The quality issues also raise concerns about students being misled into thinking their testing is more thorough than it actually might be.

\paragraph{Lack of Trust}
A related but distinct concern was a \textbf{lack of trust} in ChatGPT’s output (9 mentions). Even when the tool produced runnable test cases, participants often second-guessed the correctness or coverage of the results. Key issues included:
\begin{itemize}
\item Inconsistencies in generated outputs across similar prompts.
\item Test cases that passed trivially without real validation.
\item Redundant or irrelevant tests that added noise rather than insight.
\end{itemize}
This skepticism can be constructive if it encourages critical review, but it may also increase cognitive load and reduce confidence in using AI tools effectively.

\paragraph{Loss of Skill and Prevented Learning}
Some participants feared that using ChatGPT might lead to a \textbf{loss of skill} or \textbf{prevent learning} (1 mention each). These responses, while less frequent, signal concern that automation might short-circuit the learning process. When students bypass the logic-building and problem-solving steps by outsourcing them to AI, they may miss foundational concepts critical for future development tasks. This aligns with recent findings of Fan et al.~\cite{fan2025beware} that generative AI tools such as ChatGPT, while shown to be effective at improving short-term task performance, may not fully support deeper learning outcomes such as intrinsic motivation, knowledge transfer, or self-regulated learning, thus fostering dependence on technology.

\section{Discussion}
\label{section:discussion}
In this section, we interpret and contextualize our findings. We begin by examining how participants' prompting styles and strategic choices interacted, shedding light on the relationship between user control and AI assistance. We then discuss the qualitative and quantitative qualities of the resulting test suites, including test smells and assertion-level behavior. Finally, we reflect on broader themes such as control, cognitive load, and experience, and introduce a conceptual framework that captures how these factors might shape engagement in AI-assisted testing workflows.

\subsection{Interactions Between Strategy and Prompting Approach}

We categorized participants’ strategies for unit testing within an AI-assisted workflow into four types, with C1-\cone and C4-\cfour emerging as the most dominant approaches. We also examined how participants approached prompt creation for ChatGPT in the context of test generation.

Our key observation is that the primary difference across participants lies in how they obtained their test assertions: either by incrementally expanding the test suite with each prompt using an iterative approach, or all at once via oneshot generation of a complete test suite followed by refinements.

To explore whether there was a relationship between participants’ prompting approach and their overall testing strategy, we analyzed how frequently different strategies were used alongside iterative or oneshot generation. Here, strategy refers to how participants leveraged ChatGPT, ranging from heavy reliance (C1-\cone) to minimal usage with more manual implementation (C4-\cfour).

We hypothesized that participants who relied more heavily on ChatGPT (C1) would favor the oneshot generation approach, as it requires less manual involvement. This trend was supported in our data: as shown in Table~\ref{tab:combinations}, 9 out of 10 C1-dominant assignments used oneshot generation.

Moreover, participants following Strategy C4, who favored manual implementation with less AI involvement, more often used iterative prompting, which better aligns with their workflow. Two out of the three cases where C4 dominant participants used oneshot generation came from a single participant who applied it at the end of their assignment to compare results against their manually developed test suite.

\begin{table}[ht]

\centering
\resizebox{\linewidth}{!}{%
\begin{tabular}{lccccc}
\toprule
\textbf{Generation} & \textbf{C1(GPT$_{IDEA}$GPT$_{IMP}$)} & \textbf{C2(GPT$_{IDEA}$P$_{IMP}$)} & \textbf{C3(P$_{IDEA}$GPT$_{IMP}$)} & \textbf{C4(P$_{IDEA}$P$_{IMP}$)} & \textbf{Mixed} \\
\midrule
Iterative & 1 & 1 & 1 & 4 & 1 \\
Oneshot   & 9 & 0 & 0 & 3 & 4 \\
\bottomrule
\end{tabular}
}
\caption{Combinations of generation approach and strategies.}
\label{tab:combinations}
\end{table}

%These patterns suggest a gap between participants’ stated confidence \andy{First time I hear about confidence...} and their actual workflow choices. 
In interviews, 6 out of the 9 participants who used oneshot generation in at least one task said they would have been able to complete the tasks without relying on any external resources, including ChatGPT or Stack Overflow. However, in practice, many leaned toward strategies that reduced manual effort and handed off more responsibility to ChatGPT. The frequent use of oneshot generation, especially among those who relied heavily on the model, points to a preference for speed and convenience over fine-grained control. While this approach can be efficient, it also carries some risk, particularly if participants rely too heavily on AI-generated output without critically evaluating its content. 

\subsection{Code Smells and Comparative Code Quality}

Naturally, in investigating participant strategies, we also wanted to see whether certain approaches led to more successful outcomes in terms of measurable qualities of the resulting test suites. While the previous section focused on how participants worked, particularly their prompting styles and reliance on ChatGPT, this section turns to what those workflows produced.

Specifically, we evaluated each test suite based on adequacy~\cite{zhu1997} and effectiveness~\cite{petrovicICSE2021,AthanasiouTSE2014}. Our goal was to determine whether prompting methods (iterative vs. oneshot prompting) or workflow strategies (C1–C4) had a noticeable impact on the quality or completeness of the resulting tests.

In Table~\ref{tab:metrics}, we present the test suite metrics for all participants. We examined line coverage and mutation scores as indicators of test adequacy and effectiveness. However, when comparing these results across different prompting styles or strategy groups, we found no clear patterns that could distinguish one approach from another. In other words, neither a prompting method nor a strategy appeared to consistently produce better-performing test suites based on these metrics.

That said, when compared to Aniche et al.'s prior study using the same tasks to analyze how professional developers engineer test cases~\cite{AnicheTSE2022}, our participants’ test suites performed better in both line coverage and mutation scores. This suggests that having access to ChatGPT may offer a notable advantage in generating effective test cases. Nonetheless, within our own study, differences in prompting or workflow style did not translate into major differences in output quality.

To further investigate potential quality differences, we examined the presence of test smells, which are patterns in test code that may indicate maintainability or reliability issues. We used TSDetect~\cite{peruma2020tsdetect} to automatically analyze the test suites. Out of the 19 test smells the tool can detect, 7 were observed in the participants’ submissions. These smells, defined as follows in Peruma et al.~\cite{peruma2020tsdetect}, provide insight into potential shortcomings in the test suites :

\begin{itemize}
    \item \textbf{Assertion Roulette}, a test method that contains more than one assertion statement without an explanation/message (parameter in the assertion method)
    \item \textbf{Conditional Test Logic}, a test method that contains one or more control statements (i.e., if, switch, conditional expression etc.)
    \item \textbf{Eager Test}, a test method that contains multiple calls to multiple production methods
    \item \textbf{Lazy Test}, multiple test methods calling the same production method
    \item \textbf{Duplicate Assert}, a test method that contains more than one assertion statement with the same parameters
    \item \textbf{Ignored Test}, a test method or class that contains the \texttt{@Ignore} annotation
    \item \textbf{Magic Number Test}, an assertion method that contains a numeric literal as an argument
\end{itemize}

To explore whether any of the participant groupings were associated with systematically higher or lower occurrences of test smells, we conducted statistical comparisons across the identified strategies (C1 to C4) and prompting styles (iterative versus oneshot). We used the Mann–Whitney U test~\cite{mcknight2010mann} for these comparisons. However, no significant differences were found, meaning no strategy or prompting group consistently produced test suites with more or fewer test smells than the others.

%This finding suggests that for the unit testing tasks in our study, the presence of test smells does not serve as a meaningful indicator for distinguishing between different ways of working. Although participants differed in their approaches to the task, with varying levels of automation, control, and prompt structure, these variations did not result in meaningful differences in test quality as measured by the presence of test smells.

In summary, this finding suggests that for the unit testing tasks in our study, 
and provided that the participants differed in their approaches to the task, with varying levels of automation, control, and prompt structure, these variations did not result in meaningful differences in test quality as measured by the presence of test smells.

%In summary, the strategies we identified, whether characterized by greater reliance on ChatGPT or by more manual involvement, did not result in systematically \baris{higher or lower test quality.} For the tasks investigated in this study, test smells do not appear to differentiate between the resulting test suites produced by different workflows.

\subsection{Average Time to Write an Assertion}

\begin{figure}[!t]
    \centering
    \includegraphics[width=\linewidth]{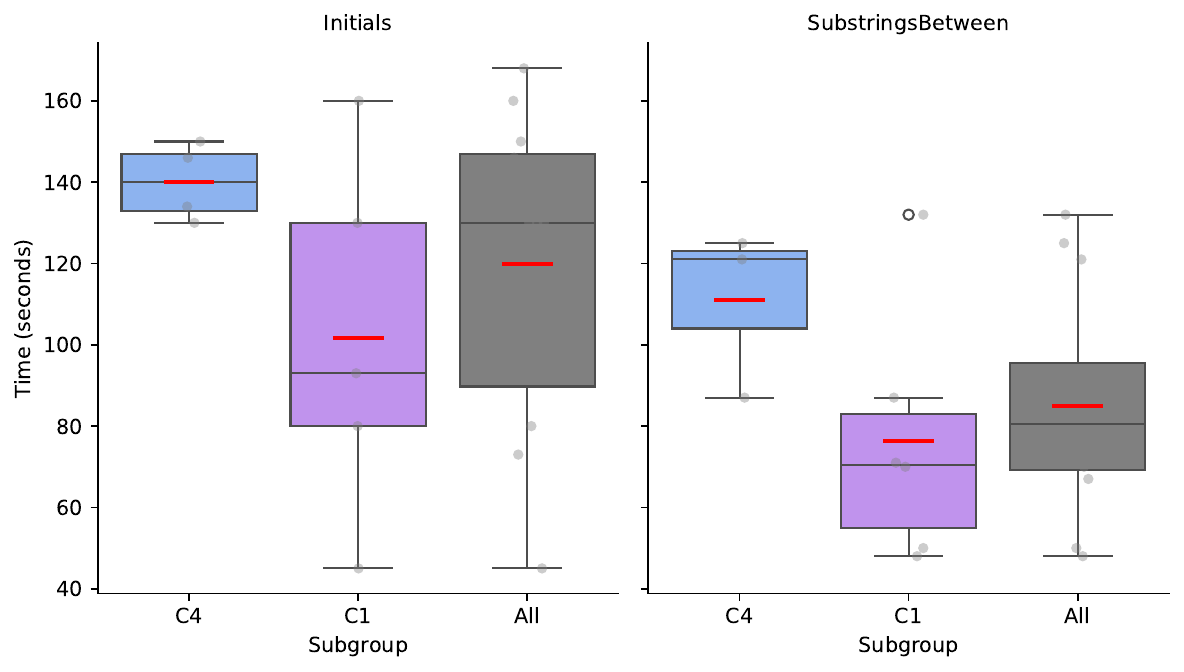}

    \caption{Assertion completion time across assignments, grouped by participant strategy. The lines in red are average values (seconds)}
    \label{fig:timeToAssertion}
\end{figure}

Figure~\ref{fig:timeToAssertion} visualizes the time participants took to complete a single assertion for each assignment. While participants who relied more heavily on ChatGPT-assisted workflows, particularly those who followed the \cone strategy, tended to complete assertions more quickly on average, the difference was not substantial enough to support the claim that ChatGPT use reduces the completion time for this task.

This suggests that the main benefit of ChatGPT-assisted workflows may lie more in reducing the cognitive or manual effort required, rather than providing a significant time advantage. The tasks, although not overly complex, are not trivial either and still require a level of reasoning and correction that ChatGPT cannot fully automate.

A notable pattern among ChatGPT reliant participants was the tendency to adopt a ``oneshot'' generation strategy, attempting to produce an entire test suite, including assertions, in a single prompt. While this approach could initially reduce the time spent writing, it often led to errors. We observed that all participants who used oneshot generation needed to manually modify the generated test cases. The majority of these were simple fixes, such as changing incorrect expected values in the generated assertions. Participants subsequently spent time identifying and correcting these issues, which decreased the net time savings.

\begin{lstlisting}[style=javaCustom, caption={Example of GPT-generated test with incorrect assertion}]
/*
 * Test case to verify extraction of initial characters from a string
 * containing multiple words separated by a specific delimiter.
 */
@Test
public void testMultipleWordsWithCustomDelimiter() {
    assertEquals("CST", initials("Customer: Service: Team", ':'));
}
\end{lstlisting}

\noindent\textit{Corrected version of the assertion:}
\begin{lstlisting}[style=javaCustom]
assertEquals("C  ", initials("Customer: Service: Team", ':'));
\end{lstlisting}

In this example, ChatGPT incorrectly assumes that each capitalized word contributes its first character, overlooking that the \texttt{initials} method only takes the first character of the original string and the segments after the custom delimiter. As ``Service'' and ``Team'' follow colons but begin with whitespace, their initials result in two space characters. This subtle misunderstanding exemplifies the most common type of assertion error produced by ChatGPT. Assignment implementations are provided in our data package~\cite{datapackage}.

It is likely that the frequency and nature of such mistakes would vary across different tasks, making our findings on time savings difficult to generalize beyond this context. Nonetheless, these subtle assertion errors represent the primary barrier to fully automating test generation with ChatGPT in our setting. At the same time, our analysis suggests that the test ideas proposed by ChatGPT were generally sound, with participants frequently choosing to incorporate them into their final test suites. This indicates that while the implementation details often required correction or refinement, the conceptual value of ChatGPT’s suggestions was high. As the capabilities of generative models continue to improve, it is plausible that these implementation errors will become less frequent, enabling greater automation in similar tasks. For now, however, the main advantage of using ChatGPT appears to lie in the reduced cognitive effort required to ideate and structure test cases, rather than in fully automating their execution.

\subsection{Control and Cognitive Load in ChatGPT Use}
Our study, focused on students as a relatively low-experience user group, investigates how individuals navigate ChatGPT-assisted workflows in software testing contexts. A key observation is that participants often opted to \textit{yield control} to the tool, likely as a means of reducing \textit{cognitive load}. Rather than extensively structuring ChatGPT inputs or outputs, they leveraged the tool to generate substantial portions of the task. This behavior appeared to be effective in our context, where tasks were of \textit{lower complexity}, and thus the risks associated with reduced user control were also lower.

However, control was not entirely absent. We identified naturally occurring strategies that helped participants \textit{retain partial control} over the task without fully reasserting manual effort. Two notable prompting strategies, labeled as \textbf{C2} and \textbf{C3}, exemplify this. In \textbf{C3-\cthree}, participants provided the \textit{test logic} and prompted ChatGPT to generate a fitting implementation. In \textbf{C2-\ctwo}, they asked ChatGPT for ideas of test cases and they then supplied the \textit{implementation}. These approaches allowed users to offload parts of the task while anchoring the generation process in their own contributions, maintaining oversight and direction.

A further method to preserve control was the use of \textit{iterative prompting} rather than oneshot generation. This strategy gave participants the ability to course-correct, reflect, and adjust ChatGPT outputs incrementally. %, rather than relying on a single, high-stakes generation. 
Though probably more cognitively demanding than oneshot use, it served as a control mechanism that aligned well with participants' cautious engagement with the tool.

While our tasks were intentionally limited in complexity, we can speculate that as \textit{task complexity increases}, these lightweight control mechanisms may become insufficient. More complex problems not only introduce a higher cognitive burden for users, but may also push the boundaries of what ChatGPT can handle effectively. In such settings, we speculate that users, especially those with limited experience, would need to exert more deliberate and structured control over the tool to ensure correctness, coherence, and alignment with task goals. Lower experience can also lead to lower rigor for correctness. Even before the advent of generative AI technologies that promise to automate testing tasks, students struggled to grasp the value of testing \cite{pham2014enablers}. As generative AI automation becomes more widespread, being informed on why we test and what correctly done testing looks like becomes even more critical. %\baris{TODO Here we have the opportunity to plug in education-related improvements that can be done, maybe} \andy{Yes, good thinking. We could argue for the necessity to be critical towards LLM code results?}

Recent findings on novice programming with generative AI by Prather et al.~\cite{prather2024widening} support this by suggesting that less experienced or lower self-efficacy students may struggle to engage productively with GenAI tools, sometimes developing misconceptions or an illusion of competence without realizing it. %Generative AI usage experiences of software engineering students by Choudhuri et al.~\cite{choudhuri2024insights} also support this by 

\subsection{A Broader Perspective}

Although our findings are rooted in a low-experience sample working on relatively simple tasks, they contribute to a broader understanding of \textit{how control is negotiated} in human-AI collaboration. We hypothesize that \textit{experience} and \textit{task complexity} are two key dimensions shaping the degree and nature of control required. While our participants leaned toward yielding control under manageable conditions, it remains an open question how this dynamic evolves as experience increases or task complexity rises.

Future research should examine how higher-experience users balance control and delegation, and whether similar strategies (e.g., C2, C3, iterative prompting) persist or evolve. Likewise, studies involving more complex, real-world programming tasks can test whether the informal control mechanisms observed here remain viable or require augmentation through information outside of users' personal experiences, such as training on AI agent use in software testing .

By mapping out these relationships, we move closer to a fuller understanding of how control in generative AI-assisted workflows can support effective and responsible \textit{software testing practices}.

%\subsubsection{Summary}

In our study, students from a low-experience group tended to yield control to reduce cognitive load on lower-complexity tasks. Nevertheless, we observed naturally occurring strategies (C2-\ctwo and C3-\cthree) and \textbf{iterative prompting} that helped preserve partial control over the process.

\begin{figure}[ht]
\centering
{\small
\begin{tikzpicture}

  \fill[green!20] (0,0) rectangle (3.25,3.25);
  \fill[yellow!20] (3.25,0) rectangle (9.5,3.25); % Bottom-right
  \fill[yellow!20] (0,3.25) rectangle (3.25,6.5); % Top-left
  \fill[yellow!20] (3.25,3.25) rectangle (9.5,6.5); % Top-right
  % Axes
  \draw[->] (0,0) -- (9.5,0) node[anchor=north] {Experience};
  \draw[->] (0,0) -- (0,6.5) node[anchor=south] {Task Complexity};

  % Grid lines
  \draw[dashed] (3.25,0) -- (3.25,6.5);
  \draw[dashed] (0,3.25) -- (9.5,3.25);

  % Quadrant labels
  \node at (1.6,5) {\textbf{C2 / C3 / Iterative}};
  \node at (6.5,5) {\textbf{Iterative prompting
}};
\node at (6.5,4.5) {\textbf{Partial oneshot
}};
  \node at (1.6,1.6) {\textbf{Most things work}};
  \node at (6.5,1.6) {\textbf{Oneshot / Minimal intervention
}};

  % Axis labels
  \node[rotate=90] at (-0.6,1.6) {Low};
  \node[rotate=90] at (-0.6,5.1) {High};
  \node at (3.25,-0.6) {Low \hspace{3.5cm} High};

\end{tikzpicture}
}
\caption{Conceptual framework illustrating how experience and task complexity might shape the recommended strategies in AI-assisted software testing. The green quadrant is the scope of this study, while the yellow quadrants reflect our current hypothesis of what can be recommended. %\andy{Should we indicate in color what we have studied, and what still needs to be investigated? While it is mentioned, we should be ultra careful not to overstate}
}
\label{fig:control-framework}
\end{figure}
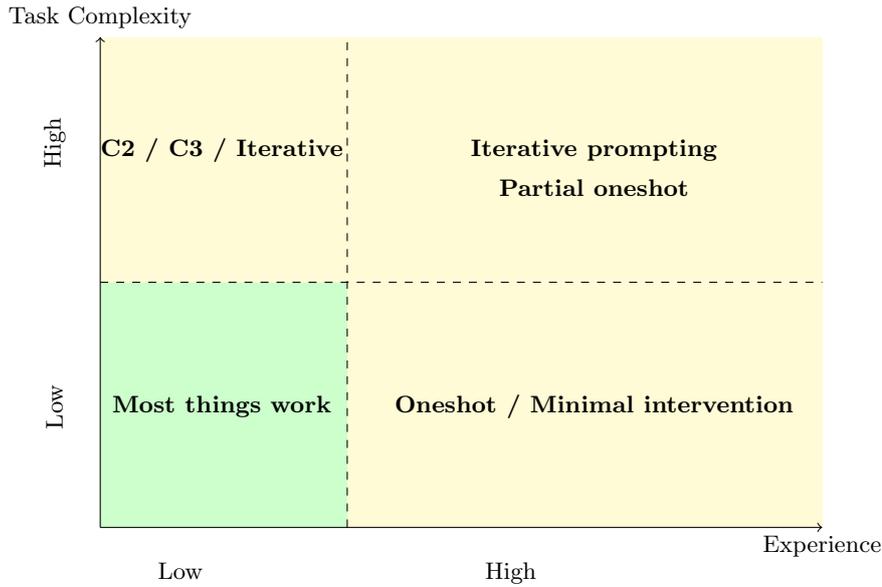

We propose a conceptual framework (Figure~\ref{fig:control-framework}) that organizes control strategies across two key dimensions: user \textit{experience} and \textit{task complexity}. This model illustrates how different combinations of these factors may influence the selection and viability of prompting strategies in software testing.

This framework provides a foundation for understanding how users might balance cognitive effort and oversight when working with generative tools. While grounded in low-experience users and low-complexity tasks, it opens space for future research to explore how control strategies evolve across different levels of expertise and problem difficulty. Ultimately, a more complete theory of generative AI-assisted \textit{software testing} will require empirical studies with broader user groups and more complex, real-world scenarios.

\paragraph{Interpretation of the Framework:}

\begin{itemize}
    \item \textbf{Low Experience, Low Complexity}: Users often yield control, relying on oneshot prompting. C2 and C3 strategies, alongside iterative prompting, can still emerge naturally to retain partial control over the process. However, lack of control related drawbacks are not observable due to low complexity.
    
    \item \textbf{Low Experience, High Complexity}: Users need to exercise more control to avoid negative outcomes. Iterative prompting and control-preserving strategies become more important to enable critical evaluation and adjustment of ChatGPT outputs.
    
    \item \textbf{High Experience, Low Complexity}: Users can effectively leverage oneshot prompting or minimal intervention strategies due to their stronger evaluation skills and domain expertise.
    
    \item \textbf{High Experience, High Complexity}: Even experienced users may shift toward iterative or reflective workflows to manage the increased likelihood of model errors or misalignment with task goals.
\end{itemize}

\section{Threats to Validity}
\label{section:threats}
We reflect on potential threats to the validity of our findings, dividing them into internal and external concerns. Where relevant, we outline measures taken to mitigate these risks.

\subsection{Internal Validity}

Two potential threats are related to \textit{participant selection and task sequencing}. All participants were undergraduate students from a single institution who had completed a mandatory software testing course. While this ensured a minimal knowledge baseline, it limits the diversity of perspectives and experiences, introducing a selection bias. Additionally, participants completed two tasks in a fixed order during a single session. This raises the possibility of maturation effects, where experience gained in the first task influenced behavior in the second, although we did not counterbalance the task order to address this.

A second category of threats concerns \textit{the influence of observation and data collection methods}. The think-aloud protocol, observation notes, and post-task interviews may have altered participant behavior. To mitigate this, we communicated to participants that the study focused on their usage patterns rather than performance outcomes. Additionally, the researcher conducting the observations had no involvement in teaching or assessment activities related to the participants. We also conducted preliminary pilot sessions to refine our procedures and familiarize both participants and researchers with the process. Moreover, we used multiple data sources, including video recordings, transcripts, and logs, to triangulate findings and reduce the impact of any one method. While the semi-structured interviews risk interviewer bias, a consistent interview guide helped maintain focus across sessions.

Finally, we acknowledge a potential threat related to \textit{data analysis and measurement}. Since our analysis was based on qualitative open coding and thematic analysis, it involves an element of subjective judgment. To mitigate this bias, we performed interrater agreement checks.

%\andy{Will remove  The next threat seems disconnected from the previous one... especially because the paragraph starts with ``finally'', it feels a bit weird. Thinking about it, this next item can also be external validity?}
%The relatively simple nature of the unit testing tasks may also limit the generality of our conclusions to more complex scenarios. However, by reusing tasks from prior research, we ensured that tasks were well-understood and also allowed meaningful comparisons.

\subsection{External Validity}

The generalizability of our results is limited by the \textit{context and characteristics of our sample and tasks}. 
Our target audience was novice testers whose testing habits and workflows had not yet been shaped by extensive experience. To represent this group, we used students who had completed a mandatory undergraduate software testing course. These students shared a common theoretical foundation and some practical exposure from the course, but lacked substantial real-world experience, making them a suitable proxy for novice testers.
These participants worked on narrowly scoped unit testing assignments. This limits ecological validity, as real-world testing often involves broader systems, messier codebases, and team dynamics. Additionally, our sample was homogenous in terms of academic background and experience, making it difficult to extrapolate findings to professional developers or to learners from different institutions.

Another key limitation is \textit{tool specificity}. Our findings are based entirely on interactions with a single version of ChatGPT. While the behaviors we observed may hold for other generative AI tools, differences in model performance or user interface could influence prompting strategies and perceived usefulness. Similarly, we focused exclusively on unit testing. Other testing practices, such as integration, regression, or exploratory testing, may yield different patterns of engagement and utility when paired with generative AI.

Lastly, we note a concern related to \textit{temporal validity}. As generative AI tools evolve rapidly, both in capabilities and user familiarity, the behaviors we observed may shift over time. However, by documenting the strategies and prompting styles in detail, we provide a snapshot that can serve as a baseline for future studies exploring how AI use in software testing matures. Additionally, the narrow scope of the assignments helps mitigate this concern, as the tasks were well within the current capabilities of AI. The AI’s effective handling of these tasks, including generating relevant test ideas, indicates that participants’ prompts were appropriately understood.

\section{Conclusion}
\label{section:conclusion}
This study examined how novice software developers interact with generative AI during unit test engineering tasks. Through an observational approach involving task recordings, the think out loud method, and post-task interviews, we investigated three key questions: what strategies do students adopt when incorporating generative AI into unit testing workflows, how they formulate prompts, and what benefits and challenges emerge from such AI-assisted workflows.

Our findings reveal that participants employed a spectrum of strategies, ranging from full reliance on AI for both ideation and implementation (C1-~\cone), to mostly manual workflows (C4-~\cfour), with intermediate approaches (C2-~\ctwo and C3-~\cthree) enabling varying degrees of user control. Prompting styles followed a similar spectrum, with participants opting for either oneshot generation of entire test suites or iterative prompting to guide ChatGPT incrementally. These strategies were not arbitrary: participants tended to match prompting style to their overall approach, participants who relied more heavily on ChatGPT for both ideation and implementation (C1-~\cone) were more likely to favor oneshot prompting. At the same time, manual or control-preserving workflows leaned more on iterative engagement.

Students reported benefits, particularly in terms of time-saving and reduced cognitive load, suggesting that generative AI can streamline repetitive aspects of test creation and help shift attention to more analytical parts of the task. However, several drawbacks were also noted, including lack of trust, concerns regarding low quality of generated test cases, and a diminished sense of ownership. Notably, the most common errors were subtle assertion mismatches that required human oversight, highlighting that while ChatGPT can support ideation effectively, it falls short of fully automating high quality test implementation end-to-end.

In discussing these findings, we propose a conceptual framework linking user experience, task complexity, and control strategies. Our results suggest that low-experience users working on low-complexity tasks often cede control to ChatGPT, but still develop natural strategies (e.g., iterative prompting, partial delegation via C2-~\ctwo and C3-~\cthree) that preserve agency. As task complexity and user expertise increase, we anticipate a shift toward more structured and deliberate control mechanisms.

This framework opens several promising directions for future research. Studies involving more experienced developers or more complex testing tasks could test whether the observed strategies generalize or evolve. Additionally, future work could investigate whether the prompting strategies and control dynamics observed here extend to other testing contexts, such as integration or non-functional testing, and explore support mechanisms that help students critically evaluate generated outputs, refine partially correct suggestions, and better balance control and automation. Such efforts may contribute to establishing best practices for responsible and effective AI-assisted software testing.

% \backmatter
\backmatter

\section*{Declarations}

\textbf{Funding:} This research is sponsored by the Swiss National Science Foundation (SNSF Grant 200021M 205146), the Dutch Science Foundation NWO through the Vici ``TestShift'' project (No. VI.C.182.032).

\textbf{Ethical approval:} This study was reviewed and approved by the Human Research Ethics Committee of Delft University of Technology.

\textbf{Informed consent:} Informed consent was obtained from all individual participants included in the study.

\textbf{Author contributions:} Baris Ardic: Conceptualization, Participant Recruitment, Observation, Investigation, Data Analysis, Writing-original draft. Quentin Le Dilavrec: Conceptualization, Data Analysis, Writing-original draft. Andy Zaidman: Conceptualization, Supervision, Writing-review \& editing, Funding acquisition.

\textbf{Data availability statement:} All data and analysis artifacts supporting the findings of this study are available at DOI [10.6084/m9.figshare.29429636.v1]. Personally identifiable data, such as raw interview transcripts and screen/audio recordings, are not publicly shared due to ethical and privacy considerations.

\textbf{Conflict of interest:} The authors declare that they have no conflict of interest.  
% Or specify any conflicts

\textbf{Clinical trial number:} Not applicable.

\bibliographystyle{spbasic}
\bibliography{references}% common bib file

\end{document}